\documentclass[journal abbreviation, manuscript]{copernicus}

\usepackage{multirow}
\usepackage{float}
\usepackage{booktabs}
\usepackage{commath}
\usepackage{amssymb}
\usepackage{mathtools, hyperref}
\usepackage[dvipsnames]{xcolor}
\usepackage{graphicx}
\usepackage{rotating}
\usepackage{array}
\usepackage{makecell}
\usepackage{caption}

\begin{document}
\nolinenumbers

\title{Identifying Time Patterns of  Highland and Lowland Air Temperature Trends in Italy and UK across monthly and annual scales}


\Author[1, 4]{Chalachew Muluken Liyew}{} 
\Author[2]{Elvira Di Nardo}{}
\Author[1]{Rosa Meo}{}
\Author[3][stefano.ferraris@unito.it ]{Stefano Ferraris}{} 

\affil[1]{Department of Computer Science, University of Turin, Italy}
\affil[2]{Department of Mathematics "G. Peano", University of Turin, Italy}
\affil[3]{Interuniversity Department of Regional and Urban Studies and Planning, Politecnico and Università of Turin, Italy}
\affil[4]{Faculty of Computing, Institute of Technology, Bahir Dar University, Bahir Dar, Ethiopia}



\runningtitle{TEXT}

\runningauthor{TEXT}

\received{}
\pubdiscuss{} 
\revised{}
\accepted{}
\published{}


\firstpage{1}

\maketitle

\begin{abstract}
This paper presents a statistical analysis of air temperature data from 32 stations in Italy and the UK up to 2000 m above sea level, from 2002 to 2021. The data came from both highland and lowland areas, in order to evaluate both the differences due to location, and elevation. The analysis focused on detecting trends at annual and monthly time scales, employing both ordinary least squares (OLS), robust S-estimator regression, and Mann-Kendall (MK) and Sen's slope methods. Then hierarchical clustering using Dynamic Time Warping (DTW) was applied to the monthly data to analyze the intra-annual pattern similarity of trends within and across the groups. 
\par \smallskip 
Two different regions of Europe were chosen because of the different climate and temperature trends, namely the Northern UK (smaller trends) and the North-West Italian Alps (larger trends). The main novelty of the work is to show that stations having similar locations and altitudes have similar monthly slopes by quantifying them using DTW and clustering. These results reveal the nonrandomness of different trends along the year and among different parts of Europe, with a modest influence of altitude in wintertime.
The findings revealed that group average trends were close to the NOAA values for the areas in Italy and the UK, confirming the validity of analyzing a small number of stations. More interestingly, intra-annual patterns were detected commonly at the stations of each of the groups, and clearly different between them. Confirming the different climates, most highland and lowland stations in Italy exhibit statistically significant positive trends, while in the UK, both highland and lowland stations show statistically nonsignificant negative trends. Hierarchical clustering in combination with DTW showed consistent similarity between monthly patterns of means and trends within the group of stations and inconsistent similarity between patterns across groups. The use of the twelve distance correlation matrices (dCor) (one for each month) also contributes to what is the main result of the paper, which is to clearly show the different temporal patterns in relation to location and (in some months) altitude. The anomalous behaviors detected at 3 of the 32 stations, namely Valpelline, Fossano and Aonoch Mor, can be attributed respectively to the fact that Valpelline is the lowest elevation station in its group, Fossano is the southernmost of the Italian stations, with some sublittoral influence, and Aonoch Mor has a large amount of missing values.

\par \smallskip
In conclusion, these results improve our understanding of temperature spatio-temporal dynamics in two very different regions of Europe and emphasize the importance of consistent analysis of data to assess the ongoing effects of climate change. The intra-annual time patterns of temperature trends could be also compared with climate model results. 
\end{abstract}


\newpage

\introduction  

The study of climate variability and its impact on our environment has garnered increasing attention in recent years, driven by growing concerns over the consequences of global climate change. The study of air temperature is a crucial aspect of climatology, widely examined worldwide, with the IPCC stating that warming is not observed or expected to be spatially or seasonally uniform \citep{collins2013long}. In fact, global warming is modulated by external forcing (‘signals’) and internal variability (‘noise’) \citep{li2022different}. The goal of comprehending its ever-changing nature in various regions over different time frames has many examples \citep{farooq2021annual, khavse2015statistical}. Globally, there is a consistent upward trend in air temperatures \citep{simmons2021temp_trend}. This phenomenon is not limited to global observations alone; it is also evident at regional levels, as seen e.g. throughout Europe, where air temperatures have displayed a continuous linear increase since 1985 \citep{twardosz2021warming} and in the Central Asian region \citep{farooq2021annual}. A time trend that appears to be mainly positive and reveals a significant rise in temperature was detected by ~\cite{gil2022temperature} when aggregated monthly temperature data were analyzed from $48$ contiguous US states. Furthermore, when disaggregated data on temperature anomalies were considered at the state level, a large number of states showed a significantly positive temporal trend coefficient. Remarkably, this trend turned out to include seven exceptions, all of which occurred in the Southeast. Also, $309$ stations in Canada and the United States were examined in \cite{isaac2012surface}, revealing significant warming trends, particularly in the Midwestern United States, Canadian prairies, and the western Arctic, primarily in winter and to a lesser degree in spring. A dataset from $19$ stations ranging from $1920$ to $2006$ was analyzed in  ~\cite{el2012trend}, and the result was a significantly increased trend in maximum, minimum, and average temperatures especially since 1960. The annual trend was explored in ~\cite{di2022analysis}, using the data obtained from three stations in Rome (Italy) in the period ($2000$ - $2020)$ and identified a statistically positive trend of annual mean temperature.

However, some warming hiatus occurred in the period 2004-2018 in the Northern Hemisphere, especially in Autumn and in more northern areas \citep{tang2022increasing, shen2018weak}. Before those years of hiatus, a study conducted in the United States did an interesting combined analysis of the pattern of temperature trends during the months and during the days \citep{vinnikov2002diurnal}. The analyses of that
paper were not repeated in other papers, and in our opinion, they deserve to be repeated with more recent data, in order to see if less noisy patterns can emerge.  Instead, many papers addressed the topic of the trends in the diurnal temperature range (DTR), for example \citep{shen2014spatiotemporal}. Annual and seasonal averages of DTR, maximum and minimum temperatures were considered by \cite{sayemuzzaman2015diurnal} using $249$ stations $(1950 -2010)$ in North Carolina, and the result showed a negative annual trend of diurnal temperature range and a positive trend for maximum and minimum temperatures, which were statistically significant.  The maximum temperature showed a negative trend during summer and spring, a positive trend during the autumn season, the minimum temperature showed an increasing trend in all seasons, and the diurnal temperature range showed a decreasing trend in all seasons. Notably, temperature extremes have become more frequent and intense throughout Europe in recent decades \citep{patterson2023hotdays}. According to \cite{patterson2023hotdays}, based on ERA5 reanalysis data from $1960$ to $2021,$ the hottest summer days in northwestern Europe are warming up at about twice the rate of average summer days. Additionally, the pattern is relatively unique when compared to the Northern Hemisphere. Another extensively addressed topic is altitude-dependent warming. It occurs throughout the world's highlands regions, with the Alps proving to be a notable hotspot for global warming \citep{palazzi2015elevation}. A study looking at the French Alps and adjacent areas, in neighboring Italy and Switzerland, found a clear overall trend indicating an increase of about $1^{\circ}$C in annual air temperature over 44 years, with large variations of this trend for different altitudes,
seasons, and regions. The trends are most pronounced between 1500 and 2000 m above sea level (asl)  \citep{durand2009reanalysis}. A recent paper describes in depth the physical mechanisms driving EDW in the tropics and subtropics, highlighting some drivers and, interestingly for our study, monthly variations \citep{byrne2024elevation}. Available observations suggest that Mediterranean mountains are experiencing seasonal warming rates that are largely greater than the global land average. The identification and attribution of human versus natural effects is beyond the scope of this paper. For example, a human fingerprint (Blackport et al., 2021), in the decreasing subseasonal near-surface air temperature variability has recently emerged from a reanalysis of the Northern Hemisphere extratropics. It features decreased near-surface air temperature variability over land in the high northern latitudes in autumn, further extending into mid-latitudes in winter. Therefore, using large ensembles of single-forcing model experiments, they attributed the pattern of reduced temperature variability primarily to increased anthropogenic greenhouse gas concentrations, with anthropogenic aerosols playing a secondary role.

\par \smallskip
In the literature, trends of time variables can be detected, estimated, and predicted using both parametric and nonparametric methods. Parametric methods, such as linear regression, robust regression, moving averages, or multiple regression, require validation of assumptions about the underlying distribution. For example, parametric methods were applied in \cite{vinnikov2002diurnal} to study the diurnal and seasonal cycles of trends of surface air temperature as well as in \cite{el2012trend} to quantify the seasonal and annual trends. Nonparametric methods, on the other hand, do not require assumptions about the underlying distribution, but ensure the robustness of the final conclusions. Among nonparametric methods, the most widely used are the Mann-Kendall (MK) test and Sen's slope estimator \citep{di2022analysis,mohsin2010trend,sayemuzzaman2015diurnal} since these methods are particularly suited for non-normally distributed data, even in the presence of missing values. Specifically, the MK test is used to detect the presence of trends in the investigated variables, and Sen's slope estimator estimates the magnitude of these trends. These methods have been widely used in numerous studies aimed at identifying and estimating trends in annual, seasonal and monthly temperatures in various countries and regions. All these results show that the trends of increasing monthly, annual, and seasonal temperatures are not homogeneous: in some regions, the increase was statistically significant, while in other regions statistical significance was not reached. Cluster analysis was performed by \cite{rebetez2008monthly} in Switzerland, showing a difference between low and high-elevation stations.

\par \smallskip

In this work, we considered a limited number of stations (32) for the sake of clarity of the proposed method of analysis. Regarding the limited number of years (20), we wanted to limit inhomogeneities in instrumentation, land use, and nonlinearity of trends. The nonlinear behavior of the last decades is also confirmed by the fact that Brunetti et al. (2006) have shown different Italian historical station trend results only adding eight more recent years in the series, in comparison with their previous analysis. In order to compare different areas and different altitudes, the attention is focused on six groups of stations over the period 2002 - 2021: eleven Italian highland stations (IH), twelve Italian stations at low altitudes (IL), five UK highland stations (UKH) and four low altitudes ones in the UK (UKL). Italian stations (both lowlands and highlands) were further stratified by distinguishing between those in the Valle d'Aosta and those in Piemonte for a total of 6 different regions.  The highland stations are between 1029 and 2017 m asl in Italy and between 773 and 1237 m asl in the UK. The lowland stations are between 232 and 577 m asl in Italy and between 140 and 249 m asl in the UK. Temperature trends are analyzed at annual and monthly time scales. 
   
\par \smallskip
The objective of this study is preliminarily to assess trends in six Italian and UK groups of stations, examining the differences between parametric and nonparametric methods in quantifying air temperature
trends, and exploring the implications of these different methods. For this purpose, Shapiro’s test is applied to test the hypothesis
of the normal distribution when necessary, as suggested by Royston (1982). After that, the main objective is to analyze the intra-annual pattern similarity of trends within and across these groups of stations, in order to assess the role of elevation and of geographical location, at small and large distances. Regarding elevation, three of them are in the highlands region, and three are in the lowlands region. Regarding the geographical location, two are from the UK, a region with low time variation of temperature, and four are in the Alps, a hotspot of global warming. Also, two different subregions in the Alps are considered in order to see the effect of small geographical distance, with respect to the long distance between the UK and Italy. Finally, hierarchical clustering and distance correlation are used to identify pattern similarity.

\par \smallskip
The paper is organized as follows: the dataset is presented in Section~\ref{sec: material} along with a brief summary of the methods employed for the analysis. Results and discussion are reported in Section ~\ref{sec: result}. Some concluding remarks are given in the last section.

\section{Material and methods} \label{sec: material}
\subsection{Study area and dataset} \label{sub:dataset}

\begin{figure}[H]
   \centering
    \includegraphics[width=\textwidth]{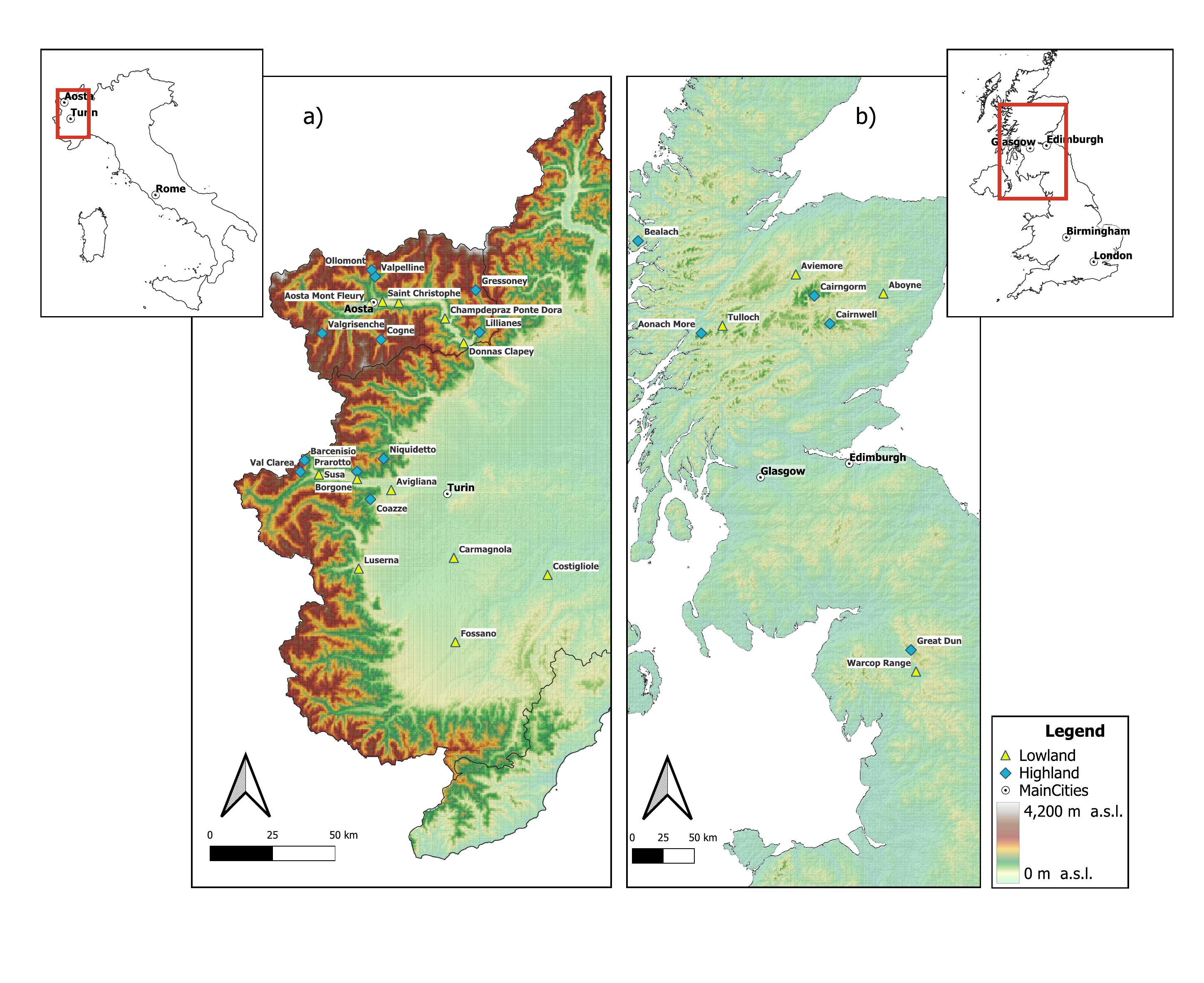}
    \caption{The geographical} distribution of all stations of (a) Italian highland and lowland, and (b) UK highland and lowland considered in this study.
    \label{fig:geographical_location}
\end{figure}

This study uses air temperature data obtained from 32 stations located in different geographical areas.  The dataset includes observations from highland and lowland stations in the UK and Italy. Specifically, five UK highland stations, four UK lowland stations, eleven Italian highland stations, and twelve Italian lowland stations were chosen to examine and analyze monthly and intra-annual patterns of air temperature trends. The geographical area locations of the study are shown in Figure ~\ref{fig:geographical_location} 
    
These observations cover the time frame from 2002 to 2021. Annual and monthly trends are calculated only over 20-year periods because of accelerated warming in the Alpine region \citep{mudelsee2019trend}, especially where trends are larger. Also, in the 1990s, measurements shifted from mechanical instruments in shelters to small electronic sensors. At all of these stations, temperature records were collected at half-hourly intervals, totaling $350640$ records for the Italian lowlands, and at one-hourly intervals, totaling $175320$ records for the highland stations in Italy and the UK as well as all lowland stations in the UK. The location and altitudes of each station in the dataset are given in Table~\ref{tab:dataset-distribution-location}. 

\par \smallskip

There are instances of missing values in the dataset, and their proportions in relation to each station can be found in the last column of Table \ref{tab:dataset-distribution-location}. The occurrence of missing values at Italian stations is minimal, while highland stations in the UK have a relatively higher percentage of missing values. Despite these variations, the total number of air temperature observations is sufficiently large, and the limited presence of missing values and their random occurrence in the dataset ensures that their influence on the analysis remains marginal. Given the small number of missing values, the classic seasonally segmented missing-value imputation technique\footnote{The function {\tt R} {\tt na\_seasplit()} of the package {\tt imputeTS} (version 0.3) was used to restore the missing information ~\citep{imputeTS}.} was employed.  When used as a pre-processing step, this method involves segmenting the time series into seasonal blocks, after which imputation is performed individually for each block using interpolation algorithms. After incorporation of the imputed values, the dataset was further processed into monthly and annual time series, as presented in the following section. 
\setcounter{table}{0}
\begin{table}[H]
    \centering
    \resizebox{18cm}{10cm}{
    \begin{tabular}{l|l|l|c|c|c|c}
        \toprule
        \toprule
        {\footnotesize \bf{Group}} & {\footnotesize \bf{Region}} & {\footnotesize \bf{Station location}} & {\footnotesize \bf{Latitude}} & {\footnotesize \bf{Longitude}} & {\footnotesize \bf{Altitude (in meter)}} & \bf{$\%$} {\footnotesize \bf{missing values}} \\
        \midrule
        
        \multicolumn{2}{c|}{\multirow{5}{*}{\parbox{2cm}{UK \\ highland }}} &
        {\footnotesize Cairngorm (CR)} & $57.0607$ °N & $-3.6066$ °E & $1237$ & $10.94$ \\
        \multicolumn{2}{c|}{} & {\footnotesize Aonach Mor (AN)} & $56.8168$ °N & $-4.9603$ °E & $1130$ & $23.33$ \\         
        \multicolumn{2}{c|}{} & {\footnotesize Cairnwell (CW)} & $56.8793$ °N & $-3.4213$ °E & $928$ & $13.60$ \\
        \multicolumn{2}{c|}{} & {\footnotesize GreatDun (GD)} & $54.6833$ °N & $-2.4500$ °E & $847$ & $7.74$ \\
         \multicolumn{2}{c|}{} & {\footnotesize Bealach (BL)} & $57.4167$ °N & $-5.7167$ °E & $773$ & $3.88$ \\ 

         \midrule
                
         \multicolumn{2}{c|}{\multirow{5}{*}{\parbox{2cm}{ UK \\ lowland }}} & {\footnotesize Aviemore (AR)} & $57.2005$ °N & $-3.8282$ °E & $228$ & $0.58$ \\
        \multicolumn{2}{c|}{} & {\footnotesize Aboyne (AY)} & $57.0767$ °N & $-2.7803$ °E & $140$ & $1.46$ \\
        \multicolumn{2}{c|}{} & {\footnotesize Tulloch (TH)} & $56.8667$ °N & $-4.7067$ °E & $249$ & $0.29$ \\
        \multicolumn{2}{c|}{} & {\footnotesize WarcopRange (WR)} & $54.5344$ °N & $-2.3900$ °E & $227$ & $0.85$ \\

         \midrule

       \multirow{12}{*}{\parbox{2cm}{Italian\\lowland }}
        & & {\footnotesize Saint Christophe (SC)} & $45.7393$ °N & $7.3634$ °E & $545$ & $0.58$ \\
        & & {\footnotesize Champdepraz Ponte Dora (CP)} & $45.6818$ °N & $7.6737$ °E & $370$ & $1.7$ \\
         & Valle d'Aosta & {\footnotesize Marcel Surpion (MS)} & $45.7366$ °N & $7.4446$ °E & $540$ & $0.75$ \\
         & & {\footnotesize Donnas Clapey (DC)} & $45.5966$ °N & $7.7664$ °E & $318$ & $0.58$ \\
        & & {\footnotesize Aosta Mont Fleury (MF)} & $45.7305$ °N & $7.2990$ °E & $577$ & $0.32$ \\

        \cmidrule{2-7}
        &  & {\footnotesize Luserna (LS)} & $44.80844$ °N & $7.24601$ °E & $475$ & $0.08$ \\
        & & {\footnotesize Susa (SS)} & $45.1386$ °N & $7.0484$ °E & $470$ & $0.09$ \\
         & & {\footnotesize Costigliole (CT)} & $44.7866$ °N & $8.1822$ °E & $440$ & $0.03$ \\
        & Piemonte& {\footnotesize Fossano (FS)} & $44.5496$ °N & $7.7251$ °E & $403$ & $0.06$ \\
         & & {\footnotesize Borgone (BG)} & $45.1229$ °N & $7.2380$ °E & $400$ & $0.22$ \\
         
        & & {\footnotesize Avigliana (AG)} & $45.0841$ °N & $7.4071$ °E & $340$ & $0.12$ \\
       
        & & {\footnotesize Carmagnola (CM)} & $44.8462$ °N & $7.7177$ °E & $232$ & $0.16$ \\   
                  
        \bottomrule
        \multirow{11}{*}{\parbox{2cm}{Italian\\highland  }}  & & {\footnotesize Valclarea (VC)} & $45.1477$ °N & $6.9567$ °E & $1068$ & $0.11$ \\
         & & {\footnotesize Prarotto (PR)} & $45.1490$ °N & $7.2370$ °E & $1431$ & $0.14$ \\
        & Piemonte & {\footnotesize Niquidetto (NI)} & $45.1937$ °N & $7.3692$ °E & $1416$ & $0.10$ \\
        & & {\footnotesize Coazze (CO)} & $45.0515$ °N & $7.3039$ °E & $1130$ & $0.15$ \\
        & & {\footnotesize Barcenisio (BA)} & $45.188$ °N & $6.9774$ °E & $1525$ & $0.79$ \\
        \cmidrule{2-7}
        
        & & {\footnotesize Gressoney (GR)} & $45.7796$ °N & $7.8258$ °E & $1642$ & $0.29$ \\
        & & {\footnotesize Cogne (CG)} & $45.6083$ °N & $7.3561$ °E & $1682$ & $0.48$ \\
        & & {\footnotesize Valgrisenche (VG)} & $45.6297$ °N & $7.0640$ °E & $1859$ & $0.24$ \\
        & Valle d'Aosta & {\footnotesize Ollomont (OL)} & $45.8494$ °N & $7.3102$ °E & $2017$ & $0.42$ \\
        & & {\footnotesize Lillianes (LL)} & $45.6337$ °N & $7.8442$ °E & $1256$ & $0.13$ \\        
        & & {\footnotesize Valpelline (VP)} & $45.8263$ °N & $7.3273$ °E & $1029$ & $0.34$ \\      
    \bottomrule
        
        \bottomrule
    \end{tabular}
    }
    \vspace{0.02em} 
    \caption{{\small Location, latitude, longitude, altitude (in meters), and missing value percentage of the weather stations where the air temperature was registered.}}
    \label{tab:dataset-distribution-location}
\end{table}

\subsection{Methods} \label{sub: method}
\subsubsection{Parametric and nonparametric methods for temperature trends.} \label{sub: method_trend}
In this section, we shall briefly recall the parametric and nonparametric methods used to detect and quantify monthly or annual mean temperature trends as set out in the second part.

For linear regression\footnote{The {\tt R} function {\tt lm()} was used.}, the monthly and annual mean temperature $y_{t}$ (in degrees Celsius)  is regressed on the explanatory variable $t$ (month or year respectively), that is $y_{t}= \beta \, t  + \epsilon,$  see for example \cite{hyndman2018forecasting}. 
Positive values of the slope $\beta$ show increasing trends, while negative values indicate decreasing trends. The coefficient of determination $R^2$ measures how much the temperature variability is attributable to the time period. Usually, if the residuals are independent and normally distributed around zero, a classical hypothesis test assesses a significant trend if the null hypothesis $\beta = 0$ is rejected at the $0.05$ level 
 \citep{wooldridge2000basic}.  A widespread method to compute an estimate of $\beta$ is the ordinary least-squares (OLS) procedure.  However, this method has a twofold drawback. First, the hypothesis of the normal distribution of residuals needs to be validated. Secondly, a single outlier can have a significant effect on the estimation to the point of invalidating the trend interpretation \citep{rousseeuw1984least}. 

On the other hand, the effect of outliers is tolerated by the robust regression\footnote{The  {\tt R} 
function {\tt lmrob()}  of the package {\tt robustbase} (version 0.99-0) was used  \citep{lmrob}.}, which allows a different distribution of residuals, \citep{rousseeuw1984robust}.
In the dataset considered here, the hypothesis of normal distribution is also sometimes violated due to the presence of outliers (see Figure~\ref{fig:annual_distribution}). Therefore, a robust regression procedure was applied to assess the temperature trends of the 32 stations. As before, a $p$-value less than $0.05$ assesses an estimated slope $\beta$ significantly different from zero. There are various methods to estimate the slope robustly. In this paper, the estimation was carried out using the so-called $s$-estimator. Suppose $(t_1, y_1), \ldots, (t_n, y_n)$ is the sample dataset. Let $\rho$ be a symmetric, continuously differentiable function with $ \rho(0)=0,$ such that $\rho$ is strictly increasing on $[0,c]$ and constant on $[c,\infty),$ with $c$ a suitable positive constant.
Suppose $f(x)$ is the standard normal probability density function and set 
$k = \int_{-\infty}^{\infty} \rho(x) f(x) {\rm d}x.$ The $s$-estimator of $\beta$ is 
\begin{equation}
\widehat{\beta} = \hbox{\rm argmin}_{\beta} s[r_1(\beta),\ldots,r_n(\beta)]
\label{(2)}
\end{equation}
with $r_i(\beta)=y_i- \beta t_i$ and $s[r_1,\ldots,r_n]$ the solution of 
\begin{equation}
\frac{1}{n} \sum_{i=1}^n \rho \left( \frac{r_i}{s} \right) = k.
\label{(3)}
\end{equation}

The results obtained from the previous methods have been further verified by
using the MK test and Sen's slope estimator method. The MK test is one of the most widely used nonparametric methods to detect trends in time series, having applications in different fields of research such as hydrology and climatology \citep{radhakrishnan2017climate}. The magnitude of the trend is usually measured by Sen's slope estimator \citep{bhuyan2018trend, radhakrishnan2017climate}. Both these nonparametric methods might be appropriately used for non-normally distributed censored time series including missing values. In the MK test\footnote{The {\tt R} function {\tt mk.test()}  of the package {\tt trend} (version 1.1.5)   was used \citep{trend}.}
the following assumptions hold: {\it i)} in the absence of a trend, observations are independent and identically distributed, that is, the observations are not serially correlated over time; {\it ii)} observations are representative of actual conditions at the time of sampling; {\it iii)} sample collection, management, and measurement methods provide unbiased and representative observations of underlying populations over time. Therefore, for the two-sided test, the zero hypothesis is that the time series has no monotonic trend. If $N$ is the sample size, the MK test statistic is calculated according to \citet{mann1945nonparametric}
\begin{equation}
    S=\sum_{i=1}^{N-1} \sum_{j=i+1}^{N} sgn(X_{j} - X_{i}), 
 \end{equation}
where $sgn$ is an indicator function taking values $-1,1$ or $0$ according to its negative, positive, or equal to $0$ (tie) argument. Thus, the MK statistics returns the sum of the number of positive differences minus the number of negative differences for all the considered differences. Note that $E[S]=0$ and  the variance  including the correction term for ties is
 \begin{equation}
 {\rm Var}(S) = \left[ \frac{N(N-1)(2N+5) - \sum_{k=1}^{n} t_{k}(t_{k}-1)(2t_{k}+5)}{18} \right] 
 \end{equation}
where $n$ is the number of tied groups and $t_{k}$ is the size of the $k^{th}$ tied group. The statistic $S$ is approximately normally distributed, with
score
$$Z= \left\{ \begin{array}{cc} 
\frac{S-1}{\sqrt{{\rm Var}(S)}}, & \hbox{\rm if $S >0$} \\
0,  & \hbox{\rm if $S =0$} \\
\frac{S+1}{\sqrt{{\rm Var}(S)}}, & \hbox{\rm if $S <0.$} 
\end{array} \right.$$

If the $p$-value of the test is below an appropriate significance level ($0.05$ and $0.01$), then there is statistically significant evidence of the presence of a trend in the time series data.  Before applying the MK test, the data were tested for serial correlation which can severely affect the results, and confirmed that there was no serial correlation in the annual and monthly mean temperature. Indeed, positive auto-correlation among the data would increase the chances of rejecting the null hypothesis, even if there is the absence of a trend \citep{cox1955some}.

 The magnitude of the trend is estimated with the help of Sen's slope estimator\footnote{The {\tt R} function {\tt sens.slope()} of the package {\tt trend} (version 1.1.5)  was used \citep{trend}.} \citep{bhuyan2018trend}. The null hypothesis indicates no trend in the time series against a two-sided alternative.  Indeed, first the slope $T_{i}$ of all data pairs is  computed as:
 \begin{equation}
     T_{i} = \frac{X_{j}-X_{k}}{j - k}  \,\, {\rm for} \,\,  i= 1,2,.....,N.
 \end{equation}
Then Sen's slope estimator is calculated as the median of all slopes, that is 
\begin{equation}
    Q= \begin{cases} T_{\frac{N+1}{2}}, & \mbox{if $N$ odd} \\ 
    \frac{1}{2} \big(T_{\frac{N}{2}} + T_{\frac{N+2}{2}}\big), & \mbox{if $N$ even.}  \end{cases}
 \end{equation}
Positive values of $Q$ indicate an upward or increasing trend whereas negative values indicate a downward or decreasing trend.
 \par \smallskip

In this study, 
 nonparametric methods were used because data from some stations have non-normal distributions (see Table \ref{tab:shapiro_wilk_test} ). As discussed in Section 3.1, 
 this hypothesis fails due to the presence of outliers. 
 Therefore, we took the opportunity to assess how much the presence of outliers might influence significant trends.
  This assessment was done by applying all the previously described methods, regardless of the assumption of normal distributions, and discussing a posteriori 
 (see Section 3) the slopes of the trends and their coefficients of determination.

\subsubsection{Hierarchical clustering with Dynamic Time Warping} \label{sub:hierachical}
This section summarizes the Dynamic Time Warping (DTW)  procedure for determining the distance matrix between any two-time series and shows how Hierarchical Clustering is used to find clusters exhibiting unique patterns of behavior.

A warping path $W$ is an alignment between two sequences $X = \{x_1, x_2, \ldots, x_n\}$ and $Y = \{y_1, y_2, \ldots, y_m\},$ also with $n \ne m$, entailing a one-to-many mapping for every pair of elements.
Thus the DTW procedure is a distance measure used to measure the similarity between two time series by finding the optimal warping path between them. To this aim a distance measure is used, that is DTW looks for the optimal alignment minimizing the distance between corresponding points \citep{shen2017clustering}.  

The algorithm firstly constructs a cost matrix $C$ where each element $C(i,j)$ represents the cost of the pair $(x_i, y_j)$, determined by utilizing a distance function, such as the Manhattan
distance $d(x_i, y_j) = |x_i -y_j|$ or the Euclidean distance $d(x_i, y_j) = \sqrt{(x_i - y_j)^2},$  between two points of the time series. The Manhattan distance was chosen because it is more robust in the presence of outliers in the data, and most of the stations examined have outliers as shown in Figure \ref{fig:annual_distribution}. In contrast, the Euclidean distance amplifies the effect of outliers by squaring their differences. Additionally, the Manhattan distance is preferred over the Euclidean distance with high-size samples.  Then, a second matrix DTW is set up having the same dimension of the cost matrix. Its $(i,j)$-th element gives the distance between two sub-sequences $\{x_1,..., x_i\}$ and $\{y_1, ..., y_j\}.$ The matrix DTW is initialized as follows: DTW$(0,0) \coloneqq 0,$ as the distance between two empty sequences is $0,$ or DTW$(i,0)=$ DTW$(0,j) \coloneqq +\infty$ 
for $i>0, j>0,$ and $i \ne j,$ as no direct alignment is possible. Then the cost matrix values are calculated recursively, taking into account the following constraints on the warping paths: 
\begin{description}
\item[{\it a)}] the alignment starts at pair $(1, 1)$ and ends at
pair $(N, M);$
\item[{\it b)}]  the order of the elements in $X$ and $Y$'s path should be maintained;
\item[{\it c)}] a pair $(x_i, y_j )$ can be followed by the three possible pairs $(x_{i+1}, y_j ), (x_i, y_{j+1})$ and
$(x_{i+1}, y_{j+1}).$
\end{description}
 The recursive functions corresponding to the three possible moves are: 

 $ \rm{DTW}(i,j) = \min \begin{cases}
     {\rm DTW}(i-1, j) + wh \times C(i,j), & \text{horizontal move}\\ 
     {\rm DTW}(i, j-1) + wv \times C(i,j), & \text{vertical move}  \\
     {\rm DTW}(i-1,j-1) + wd \times C(i,j), & \text{diagonal move}  \\
 \end{cases}$ \\
 where $wh, wv,$ and $wd$ are the weights for the horizontal, vertical, and diagonal move. When all weights are equal $(wh, wv, wd) = (1, 1, 1),$ the recursive function facilitates diagonal alignment because the cost of one step is less than the cost of two steps combining the vertical and horizontal alignments. One way to balance this bias is to choose weights $(wh, wv, wd) = (1, 1, 2).$

The final DTW distance\footnote{
The {\tt R} function {\tt proxy::dist()} of the package 
{\tt proxy} (version 0.4-27) was used \citep{meyer2022package}. This distance is produced by the {\tt R} function {\tt dtwDist} of the package 
{\tt dtw} (version 1.23-1) \citep{giorgino2009computing} and registered as a distance function in the database of distances {\tt pr\_DB} of {\tt proxy}.} is the total cost of the optimal warping path which measures how well the two sequences can be aligned while minimizing the overall cost. Smaller DTW distances indicate greater similarity between the sequences, as they require less distortion to align optimally. DTW is susceptible to overfitting, which can occur, for example, if the warping window is not chosen appropriately in sequences of equal length, leading to inflated similarity scores between sequences. To overcome this drawback, a regularization technique can be introduced by adding a penalty term to the cost function aiming to penalize excessive or large warping steps.
This penalty term can be added to the original DTW cost function as follows:
${\rm DTW}_{regularized}(i,j) = {\rm DTW}(i,j) + \lambda \times  \gamma (i,j)$ where $\lambda$ is the regularization parameter, tuning the strength of the regularization, and $\gamma (i,j)$ is the regularization term. We have set $\gamma (i,j) = (i - j)^2.$ With this choice, alignment steps that have a large difference in indices are penalized, discouraging the alignment from jumping too far off the diagonal. 


  \par \smallskip
 
 Hierarchical clustering (HCA) is an algorithm for grouping similar objects into groups, 
called clusters. The distances among these objects are initially given by the regularized DTW 
distance matrix. The output is a set of clusters, where every cluster has different 
characteristics from each other, and the objects within it are broadly similar to 
one another. The algorithm\footnote{The {\tt R} function {\tt hclust()} was used. Equivalently, the {\tt R} function {\tt tsclust()} can be used from the {\tt R} package {\tt dtwclust} \citep{sarda2017comparing}.} initially
splits the sample into clusters, each containing only one sample point. A proximity matrix $D$ is initialized as $D(C_i, C_j) = {\rm DTW}_{regularized}(i, j), i \ \in X$ and $j \in Y.$ Then the two clusters having smaller proximity index are merged in a new cluster, let's say $C_{new}.$  After merging,  the proximity matrix $D$ is updated recalculating the proximity 
index between the newly formed cluster $C_{new}$ and the remaining clusters $C_k$, using the complete linkage criterion.
This criterion picks the two farthest (most dissimilar) points, such that one point lies in a cluster and the other point lies in a different cluster, and  
defines the  proximity index between these two 
clusters as the maximum regularized DTW between these two data points. The procedure continues by identifying the next pair of clusters with the smallest proximity index, merging
them, and updating the matrix
$D$ still using the complete linkage criterion. This procedure is repeated until all clusters are merged into one or until the desired number of clusters is obtained.
The final output is a hierarchical tree 
(dendrogram) that shows the sequence of merges and the distances at which 
each merge occurred. 
Although a classical way to analyze the result of HCA is to use the dendrogram, we chose a table representation (Tables \ref{tab:DTW_similar_measure} and \ref{tab:slope_cluster} for monthly mean air temperatures and their slopes respectively) to highlight which stations deviate from the geographic group they belong to. Tables \ref{tab:DTW_similar_measure} and \ref{tab:slope_cluster} show the analysis for 4 clusters that correspond to the 4 geographic areas considered. A sensitivity analysis was performed (Table~\ref{tab:silhouette_score}) that confirms the choice of the 4 clusters as the optimal choice.
\par \smallskip
Clustering performance with a given number of clusters was measured using the {\it Silhouette Score} \citep{rousseeuw1987silhouettes}. 
This index measures how similar a data point is to its cluster compared to other clusters. For this purpose, the mean intra-cluster distance $a$ is compared with the mean nearest-cluster distance $b$ for each data point.
In details, 
the mean distance between the $i$-th data point $x_i$ in $\ C_I$  and all other data points in the same cluster $C_I$ is defined as
$$a(i) = \frac{1}{|C_I| - 1} \sum_{j \in C_I, i \neq j} d(i, j)$$
where $|C_I|$ is the cardinality of the cluster and  $d(i,j)$ is the distance between $x_i$ and $x_j$ in the cluster $C_I.$ The normalization is done with respect to $|C_I| - 1$  as the distance $d(i,i)$ is not included in the sum. Therefore the smaller the value $a(i)$ is, the better  will be the assignment of $x_i$ to $C_I.$ 
Similarly, the mean dissimilarity $d(i,C_j)$ of $x_i$ to some other cluster $C_J \neq C_I$ is defined as the mean of the distance between $x_i$ and all $x_j \in C_J,$ that is
$$d(i,C_j)= \frac{1}{|C_J|} \sum_{j \in C_J} d(i, j).$$
The minimum $b(i) = \min_{J \neq I} d(i,C_j)$ of these dissimilarity indexes identifies the \lq\lq neighboring cluster\rq\rq of $x_i,$ because it is the best next-fit cluster for point $x_i$. Thus the Silhouette Score\footnote{{The {\tt R} function {\tt silhouette()} was used from the package {\tt clusters} (version 2.1.5) \citep{cluster}. }} corresponding to $x_i$ is defined as
$$
S(i) = \left\{
\begin{array}{lc} 
\displaystyle{\frac{b(i) - a(i)}{\max(a(i), b(i))}} & \text{if } |C_i| > 1, \\
0 & \text{if } |C_i| = 1.
\end{array} 
\right.
$$

An overall Silhouette Score $S$ is computed by taking the mean of all $S(i)$ values. As $S(i) \in (-1,1)$
for all data points, the same happens for $S$  ranging from $-1$ to $ 1$. Hence, a value of $S$ close to $1$ suggests that the data points are well clustered, and each one is more similar to a neighboring point in its own cluster as opposed to those within another cluster. A value of $S$ about $0$  indicates that
data points are located at or near the boundary between clusters. A negative value of $S$ is likely to suggest that data points may be better allocated in a neighboring cluster rather 
than their current cluster.

Distance correlation ~\citep{szekely2009brownian} is a dependency measure used to examine and quantify relationships between the temperature data collected for the $32$ stations. Distance correlation is not only
invariant to linear transformations but also to some nonlinear
transformations and, unlike traditional methods, does not require assumptions of normality. As with other correlation measures, distance correlation ranges from $0$ to $1$ where $0$ means no correlation and $1$ means perfect correlation. For the calculation of distance correlation \footnote{The {\tt R} function {\tt dcor.test()} of the package 
{\tt energy} (version 1.7-11) was used \citep{energy}},
suppose we have a random
sample $({\bf X},{\bf Y}) = \{(X_k, Y_k) : k = 1, \ldots , n\}$ of $n$ iid random vectors $(X,Y)$ of dimension $p$ and $q$ respectively. 
First, the Euclidean distance is
computed between different samples
$$a_{j,k} = \norm{X_{j}- X_{k}}_p \quad {\rm and} \quad b_{j,k} = \norm{Y_{j} - Y_{k}}_q \quad {\rm for} \quad \,\, j,k=1,2,3,...,n$$ 
with $\norm{.}$ is the Euclidean distance. Then, define 
$$A_{j,k} = a_{j,k} - \bar{a}_{j.} - \bar{a}_{.k} + \bar{a}_{..} \quad {\rm and} \quad B_{j,k} = b_{j,k} - \bar{b}_{j.} - \bar{b}_{.k} + \Bar{b}_{..}$$
where 
$$\bar{a}_{j.} = \frac{1}{n} \sum_{l=1}^n a_{jl} \quad \bar{a}_{.k} = \frac{1}{n} \sum_{l=1}^n a_{lk}  \quad \bar{a}_{..} = \frac{1}{n} \sum_{k,l=1} a_{kl}$$
and similarly for $\bar{b}_{j.},  \bar{b}_{.k},$ and $\bar{b}_{..}.$ These values are then used for computing the distance
covariance (dcov)
$$dcov^2(X, Y) = \frac{1}{n^2} \sum_{j=1}^{n} \sum_{k=1}^{n} A_{j,k}B_{j,k}$$
and the distance correlation (dcor)
$$ dcor^2(X,Y) = \frac{dcov^2(X,Y)}{\sqrt{dvar^2(X)dvar^2(Y)}}$$ 
where $dvar^2(X) = dcov^2(X,X)$ and $dvar^2(Y) = dcov^2(Y,Y).$

 \section{Results and discussion} \label{sec: result}
 
In the following annual and monthly trends are explored, using the parametric and nonparametric methods outlined in Section \ref{sub: method}. Hierarchical clustering, as described in Section \ref{sub:hierachical}, is employed to discover the similarity of monthly mean temperature patterns within and between each group of stations. 
The distance matrix in each month for the $32$ stations is used to find where the correlations are significant.

\subsection{Annual average temperature trends} \label{sub: annual_section} 
\begin{figure}[H]
    \centering
    \includegraphics[width=0.94\textwidth]{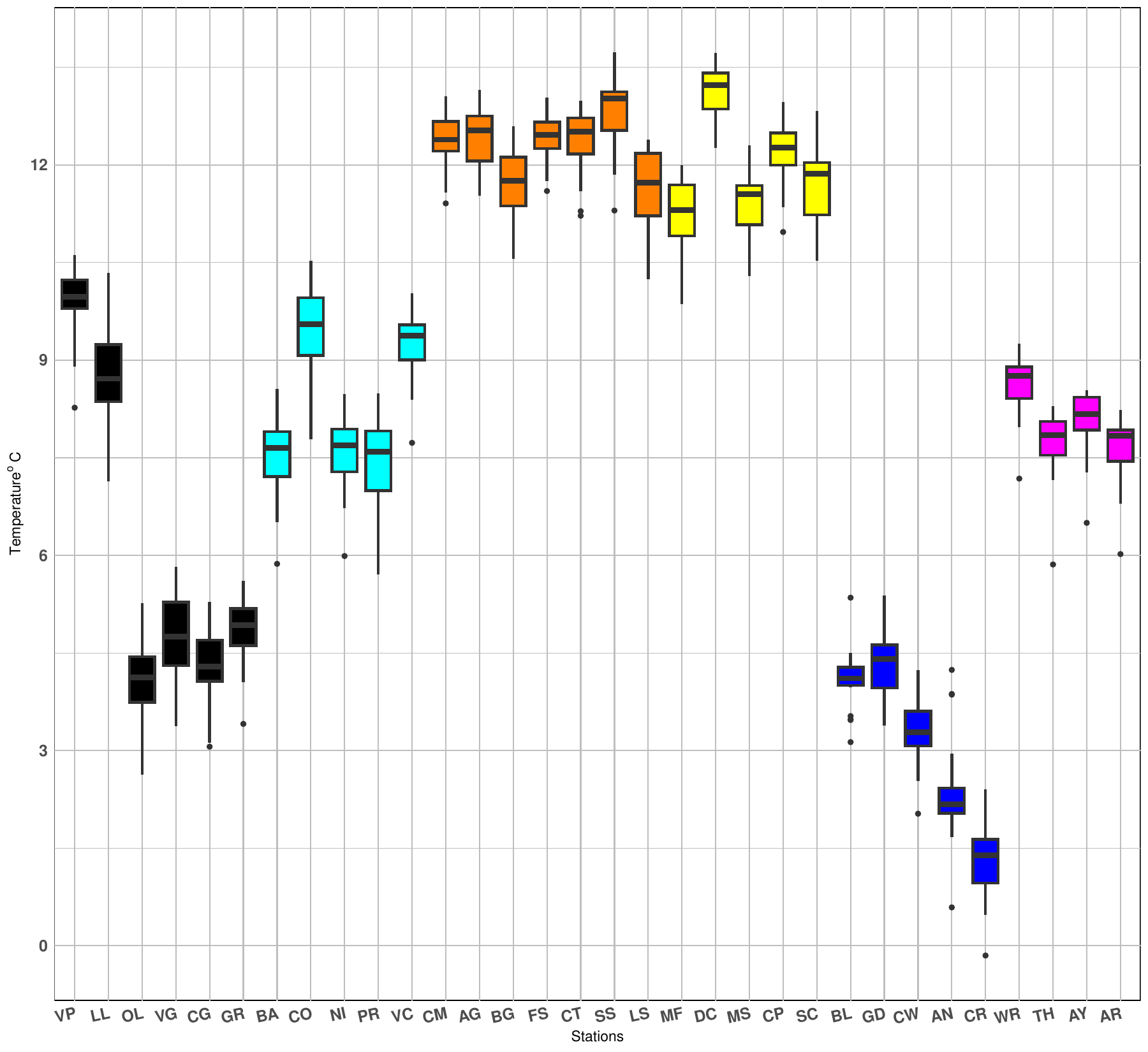}
    \caption{Box-plots of the annual average temperatures of all stations:
    black and cyan box plots represent Valle d'Aosta and Piemonte in the Italian highlands. Orange and yellow box plots correspond to Piemonte and Valle d'Aosta in the Italian lowlands, respectively, while blue and magenta box plots represent the UK highland and lowland stations.}
    \label{fig:annual_distribution}
\end{figure}

To get the annual temperatures, an averaging transformation over each year was applied, grouping half-hourly and hourly measurements into the monthly and annual time windows. For each station, box plots of the annual mean temperature time series are depicted in Figure~\ref{fig:annual_distribution}. It is evident the difference between the colder highlands (black, cyan, and blue) and the warmer lowlands (orange, yellow, and magenta), as well as between the colder UK (blue and magenta) with respect to the warmer Italian stations. In each group, some of the stations exhibit outliers, potentially affecting the hypothesis of normal distribution. Indeed a single outlier can have a substantial effect on the trends obtained by the OLS method \citep{rousseeuw1984least}. Thus, to assess this hypothesis, the Shapiro-Wilk test for each station was applied. The results of Shapiro's Wilk test are shown in Table~\ref{tab:shapiro_wilk_test}. The test rejects the null hypothesis for the stations Valpelline, Aonach Mor, Costigliole, and all UK lowland stations.
As the boxplot in Figure \ref{fig:annual_distribution} shows, all these stations have outliers. In particular, Valpelline, and all UK lowland stations exhibit a single lower outlier.  Once this outlier is removed, the Shapiro-Wilk test no longer rejects the hypothesis of normal distribution. It is worth noting that these outliers correspond to the year 2010, which was the coldest year due to the co-presence of two very cold winter months: January and December \footnote{Usually the coldest month of the seasonal cycle is either December or January.}. 
This analysis highlights the significant impact of outliers causing non-normal distribution.

\setcounter{table}{1}
\begin{table}[H]
  \centering
   \begin{tabular}{|c|c|l|l||l|l|l|}
    \toprule
    \textbf{Group} & \textbf{Region} & \multicolumn{1}{c|}{\textbf{Station}} & \multicolumn{1}{c||}{\textbf{p-value}} & \textbf{Group} & \multicolumn{1}{c|}{\textbf{Station}} & \multicolumn{1}{c|}{\textbf{p-value}} \\
    \midrule
    \multirow{12}[24]{*}{Italian Lowland} & \multirow{5}[10]{*}{Valle d'Aosta} & Mont Fleury      &  0.16     & \multirow{12}[24]{*}{UK Lowland} &  Aboyne     & {\bf 0.002}{\bf{*}} \\
&        & Donnas Clapey & 0.23   &     &WarcopRange& {\bf{0.01}{\bf{*}}} \\

&        & Marcel Surpion& 0.39   &     &  Aviemore &  {\bf 0.002}{\bf{*}}\\

&       & Champdepraz Pont Dora & 0.45 &   &Tulloch& {\bf{0.001}{\bf{*}}} \\
      
&       & Saint Christophe      & 0.52 &       &       &  \\
\cmidrule{2-4}          
& \multirow{7}[14]{*}
{Piemonte} &  Carmagnola   &  0.88              &       &       &  \\
&       &     Avigliana    &  0.32              &       &       &  \\
&       &     Borgone      &  0.65              &       &       &  \\
&       &     Fossano      &  0.33              &       &       &  \\
&       &     Costigliole  &  {\bf 0.02}{\bf{*}}&       &       &  \\
&       &     Susa         &  0.17              &       &       &  \\
&       &     Luserna      &  0.16              &       &       &  \\
    \midrule
    \multirow{11}[22]{*}{Italian Highland} 
& \multirow{5}[10]{*}{Piemonte} &  Barcenisio & 0.39 & \multirow{11}[22]{*}{UK Highland} &  Bealach    &  0.10  \\
&      &   Coazze    &  0.46 &       &  GreatDun    &  0.78 \\
&      &  Niquidetto &  0.27 &       & CairnWell    &0.79   \\
&      & Prarotto    &  0.49 &       & Aonach Mor   &  {\bf 0.01}{\bf{*}} \\
&      &  Valclarea  &  0.23 &       &  Cairngorm   & 0.63  \\
\cmidrule{2-4}         
& \multirow{6}[12]{*}{Valle d'Aosta} &  Valpelline &{\bf 0.02}{\bf{*}}       &       &       &  \\
&       & Lillianes    &  0.94  &       &       &  \\
&       & Ollomont     &  0.96  &       &       &  \\
&       & Valgrisenche &  0.80  &       &       &  \\
&       & Cogne        &  0.44  &       &       &  \\
&       & Gressoney    &  0.21  &       &       &  \\
    \bottomrule
    \end{tabular}%
    \vspace{0.1cm}
  \caption{$p$-values of Shapiro-Wilk test for normal distribution. Stations marked with bold and asterisk ${\*}$ denote statistically significant trends at the $5\%$ significance level.}
    \label{tab:shapiro_wilk_test}
\end{table}%
The normal distribution hypothesis is not rejected for all Italian highland stations except Valpelline.
It is worth mentioning that the Valpelline station is located at the lowest altitude among the Italian Valle d'Aosta highland stations. Italian lowland stations show a comparable median that is fairly symmetrical concerning
dispersion. The only exception is the Costigliole station for which data are not normally distributed. Outliers at stations in the UK highlands do not affect the distribution of annual mean temperature, except the Aonach Mor station. According to the results of the Shapiro-Wilk test for the UK lowland stations, their annual mean temperature showed a deviation from the normal distribution. The effect of this result on the annual mean temperature trend test will be examined later on.

Table~\ref{tab:annual_mean_temp_trend} shows the results of the methods outlined in subsection \ref{sub: method_trend}
to compute trends in annual mean temperatures.
The third and fourth columns refer to the OLS method and the $s$-estimator, respectively. Their slopes and $R^2$ coefficients of determination are given in each sub-column. 
The fifth column refers to the MK test and Sen's slope estimator. The subcolumns report
$p$-values, Sen's slope estimators and $R^2$ coefficients of determination, respectively. 

\setcounter{table}{2}
\begin{table}[H]
    \centering
    \resizebox{18cm}{9cm}{
    \begin{tabular}{l|l|l|c|c|c|c|c|c|c}
    \toprule \toprule
        {\footnotesize Group}& {\footnotesize Region} & {\footnotesize Station Names} & \multicolumn{2}{c| }{{\footnotesize OLS}} & \multicolumn{2}{c| }{{\footnotesize $s$-estimator}} & \multicolumn{3}{c}{{\footnotesize MK test \& Sen's estimator}} \\
         \cline{4-10}
        & & & {\footnotesize Slope} & {\footnotesize $R^2$} & {\footnotesize Slope} & {\footnotesize $R^2$}   & {\footnotesize $p$-value} & {\footnotesize Slope}& $R^2$ \\
         \midrule

\multicolumn{2}{c|}{\multirow{5}{*}{\parbox{2cm}{UK \\ highland }}} 
        & {\footnotesize Cairngorm  }  & -0.011 & 0.02 & 0.014 & 0.12 & 0.92 & -0.003& 0.04 \\
        \multicolumn{2}{c|}{} & {\footnotesize Aonach Mor}  & 0.02 & 0.02 & -0.004 & 0.01 & 0.89  & -0.002 & 0.01\\         
        \multicolumn{2}{c|}{} & {\footnotesize CairnWell } & -0.001 & 0.00   & 0.005 & 0.02 & 0.92  &-0.002& 0.00\\
        \multicolumn{2}{c|}{} & {\footnotesize GreatDun} & -0.022 & 0.06 & -0.02 & 0.20 & 0.38 &  -0.02 & 0.06 \\
         \multicolumn{2}{c|}{} & {\footnotesize Bealach} & -0.023 & 0.08 & -0.007 & 0.09  & 0.31  & -0.012 & 0.07 \\

         \midrule
                
         \multicolumn{2}{c|}{\multirow{5}{*}{\parbox{2cm}{ UK \\ lowland }}} & {\footnotesize Aviemore } & -0.007 & 0.006 & -0.01 & 0.19 & 0.54  &-0.01& 0.19\\
        \multicolumn{2}{c|}{} & {\footnotesize Aboyne} & -0.002 & 0.05 & -0.03 & 0.59  & 0.14  & -0.02 & 0.59 \\
        \multicolumn{2}{c|}{} & {\footnotesize Tulloch}  & -0.01 & 0.02 & -0.02 & 0.46 & 0.27  & -0.02 & 0.46\\
        \multicolumn{2}{c|}{} & {\footnotesize Warcop Range} & -0.002 & 0.0004 & -0.007 & 0.15 & 0.65 &  -0.003 & 0.14 \\               
        
          \midrule
         \multirow{12}{*}{\parbox{2cm}{Italian\\lowland }}
       &  & {\footnotesize Saint Christophe}  & \textbf{0.059*} & 0.34 & \textbf{0.074*} & 0.73 & \textbf{0.01*}  & \textbf{0.059*} & 0.68\\
       &   & {\footnotesize Champdepraz Ponte Dora}  & 0.038 & 0.19 & 0.036 & 0.54 & \textbf{0.02*}   & \textbf{0.038} & 0.61\\
       & Valle d’Aosta   & {\footnotesize Marcel Surpion}  & \textbf{0.052*} & 0.36 & \textbf{0.064*} & 0.79 & \textbf{0.01*}  & \textbf{0.050*} & 0.61\\
       &  & {\footnotesize Donnas Clapey}  & 0.014 & 0.04 & 0.023 & 0.27 & 0.46  & 0.011 & 0.22\\
       &   & {\footnotesize Mont Fluery}  & 0.042 & 0.16 & 0.014 & 0.14 & 0.07  & 0.046 & 0.21\\

       \cmidrule{2-10}
       &   & {\footnotesize Luserna} & \textbf{0.075*} & 0.57 & \textbf{0.084*} & 0.67 & \textbf{0.001} & \textbf{0.074} & 0.57 \\
       &    & {\footnotesize Susa} & \textbf{0.039*} & 0.22 & \textbf{0.037*} & 0.28   & \textbf{0.03*} & \textbf{0.035*} & 0.21 \\
        &   & {\footnotesize Costigliole  }  & \textbf{0.035*} & 0.22 & \textbf{0.032*} & 0.24 &\textbf{0.02*}  & \textbf{0.032*}& 0.22 \\
       & Piemonte   & {\footnotesize Fossano}  & -0.001 & 0.00 & -0.002 & 0.00 & 0.96  & -0.001 & 0.04\\
        &  & {\footnotesize Borgone } & \textbf{0.052*} & 0.41   & \textbf{0.056*} & 0.46 &\textbf{0.001*}  & \textbf{ 0.053*}& 0.41\\
        &   & {\footnotesize Avigliana} & \textbf{0.056*} & 0.53 & \textbf{0.068*} & 0.51 &\textbf{0.001*}  &  \textbf{0.061*} & 0.53 \\
        &   & {\footnotesize Carmagnola} & \textbf{0.044*} & 0.43 & \textbf{0.042*}& 0.52  &\textbf{0.002*}  & \textbf{0.043*} & 0.43 \\ 
         
        \midrule
        \multirow{6}{*}{\parbox{2cm}{Italian\\highland}}
       &  & {\footnotesize Valclarea} & \textbf{0.042*} & 0.20 & \textbf{0.040*} & 0.60 & \textbf{0.029*} & \textbf{0.040*} & 0.60 \\
       &  & {\footnotesize Prarotto} & \textbf{0.056*} & 0.24 & \textbf{0.049*} & 0.55 & \textbf{0.021*} & \textbf{0.051*} & 0.58 \\
       & Piemonte & {\footnotesize Niquidetto} & 0.039 & 0.15 & 0.018 & 0.12 & 0.098 & 0.029 & 0.19 \\
       &  & {\footnotesize Coazze} & \textbf{0.059*} & 0.28 & \textbf{0.061*} & 0.75 & \textbf{0.015*} & \textbf{0.060*} & 0.70 \\
       &  & {\footnotesize Barcenisio} & \textbf{0.053*} & 0.25 & \textbf{0.043*} & 0.66 & \textbf{0.015*} & \textbf{0.050*} & 0.67 \\
       \cmidrule{2-10}
       &  & {\footnotesize Gressoney} & 0.031 & 0.12 & 0.026 & 0.12   & 0.14 & 0.031 & 0.18 \\
       &  & {\footnotesize Cogne  }  & \textbf{0.056*} & 0.30  & \textbf{0.056*} & 0.38 & \textbf{0.02*} & \textbf{0.056*}& 0.30 \\
       & Valle d’Aosta  & {\footnotesize Valgrisenche}  & \textbf{0.068*} & 0.37 & \textbf{0.063*} & 0.41 & \textbf{0.02*} & \textbf{0.068*}& 0.37\\
       &  & {\footnotesize Ollomont } & \textbf{0.051*} & 0.23   & \textbf{0.046*} & 0.26 & \textbf{0.03*} & \textbf{ 0.048*}& 0.23\\
       &  & {\footnotesize Lillianes} & \textbf{0.089*} & 0.37 & \textbf{0.092*} & 0.36 & \textbf{0.01*} &  \textbf{0.089*} & 0.37 \\
       &  & {\footnotesize Valpelline} & -0.001 & 0.00 & 0.01 & 0.13  & 0.48 &  -0.004 & 0.01 \\
         \bottomrule \bottomrule     
    \end{tabular}
    }
    
\caption{Parametric and nonparametric methods of subsection \ref{sub: method} applied to the average annual temperatures. Stations marked with bold and asterisk ${\*}$ denote statistically significant trends at the $5\%$ significance level. The unit of slopes are ($ ^\circ C/year$).}
 \label{tab:annual_mean_temp_trend}
\end{table}

In the Italian highland group, most stations exhibit a significant positive trend, except for Valpelline, which shows a relatively small negative trend and also fails the normality test.
The Gressoney and Niquidetto stations, on the other hand, show a nonsignificant positive trend in all methods, with varying degrees of fitting, depending on the method used. 
Similarly, most of lowland stations in Italy show statistically significant positive slopes, with the exception of Fossano, which shows a nonsignificant negative trend in all methods. It is worth observing that the non-normality checked at the Costigliole station as well as the presence of outliers do not affect the statistical significance result of the three methods.

\par \smallskip

The stations in the UK highland exhibit statistically nonsignificant negative trends with the exception of the Aonach Mor which shows a positive trend in the OLS and a statistically nonsignificant negative trend with the $s$-estimator and Sen's estimator, this is because the OLS method reveals sensitivity to outliers and non-normality.
   
Lowland stations in the UK showed statistically nonsignificant negative trends; however, the robust regression and Sen's slope estimators produced slopes of similar magnitude, relatively larger than those obtained by the OLS method. This result highlights the outlier tolerance and distribution independence of robust regression and Sen's slope trend analysis. It is noteworthy that the $R^2$ value of the OLS method is smaller than the $R^2$ values of the $s$-estimator and Sen's estimator. 

\par \smallskip
As anticipated, Table \ref{tab:annual_mean_temp_trend} shows how outliers and non-normal distribution affect trend analysis. For example, UK lowland stations, with non-normal distributions (Shapiro-Wilk testm $p$ < 0.05), still have negative slopes in both parametric and nonparametric methods, indicating that outliers do not affect the trend but impact the coefficient of determination. OLS assigns less significance to trends compared to nonparametric methods. For normally distributed data, all methods detect significant trends, except for Champdepraz, where Sen’s slope found a statistically significant trend while OLS and the $s$-estimator did not. These findings confirm that Sen’s slope estimator method with the MK test is the better method for trend analysis, with or without outliers.


\par \smallskip

These results are in agreement with those found in the French Alps and some adjacent regions of Italy and Switzerland \citep{durand2009reanalysis}, where it was observed that in spite of the fluctuations in the trend varies by altitude, season, and region, there is a general trend of increasing average annual temperatures. Similar studies conducted in Spain ~\citep{el2012trend}, the Midwestern United States, the Canadian Prairies, and the Western Arctic ~\citep{isaac2012surface}, Croatia ~\citep{radhakrishnan2017climate}, and India ~\citep{bhuyan2018trend} have reported a significant warming trend in average annual temperatures, indicating an overall increase over the past century. In addition, research carried out in Gombe State ~\citep{yusuf2018trend} showed a significant increase in maximum and average temperatures, while minimum temperature showed a nonsignificant upward trend. In ~\cite{meshram2020long}, an increase in annual and seasonal temperatures between $1901$ and $2016$ is reported, focusing on Chhattisgarh State. In contrast, highland and lowland stations in the UK showed nonsignificant cooling trends, although the magnitude of the negative slope is smaller in absolute value.

\par \smallskip

\begin{figure}[H]
    \centering
    \includegraphics[width=\textwidth, height=0.6\textwidth]{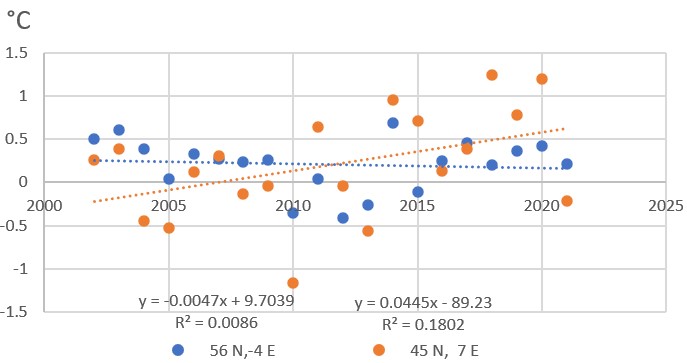}
    \caption{The anomalies of areal temperatures (with respect to 1991-2020 averages) for Italy and the UK as evaluated by NOAA (in orange the Italian and in blue the UK data). The latitude and longitude values are reported at the bottom.}
    \label{fig:anomaly}
\end{figure}

To assess the representativity of the considered $32$ stations, their trends were compared with the average trends of the corresponding Italian and UK areas. Annual anomalies with respect to the average temperature values from $1991$ to $2020$ were then calculated. The {\tt NOAA}
algorithm  ~\citep{NOAA} was used for the two (one in Italy and one in the UK) geographic areas where the stations are included. They are about $4000$ km$^2,$ as defined by the coordinates ($45$N, $7$E) and ($56$N, $-4$E), respectively. The trend slopes were very similar to the average slopes reported in Table \ref{tab:annual_mean_temp_trend}, being $0.0445^{\circ}C / year$ and $-0.0047^{\circ}C / year$ for Italy and the UK, respectively. Therefore, the here considered 32 stations provide trend values consistent with those estimated over the corresponding wide areas where they are located (see Figure \ref{fig:anomaly}).

\subsection{Monthly temperature trends} \label{sub: monthly_subsection}

The variability of monthly mean temperature trends at the $32$ stations was assessed subsequently  to an analysis of the annual mean temperatures.
To get an initial descriptive overview of the characteristics shown by the stations for monthly mean temperature trends, the MK test and Sen's estimator were applied. In the following, we briefly summarize the results highlighting similarities and differences.

\par
\begin{figure}[H]
    \centering
    \includegraphics[width=\textwidth, height=0.6\textwidth]{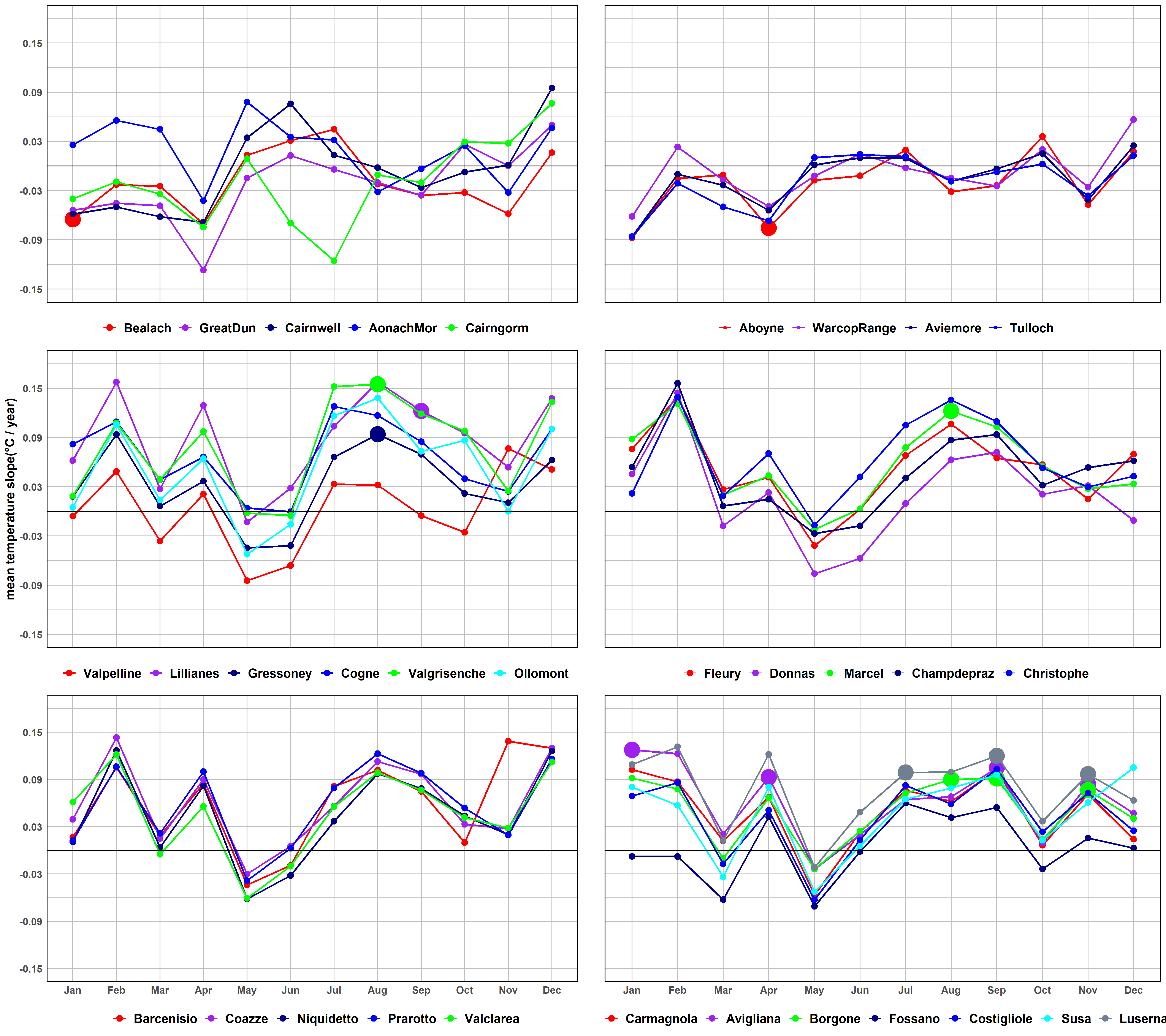}
    \caption{The trends from Sen's slope estimators of monthly temperature time series from 2002 to 2021 at the UK stations (above), and Italian ones (in the middle Valle d'Aosta stations, and below Piemonte stations). On the left are highland stations, while on the right are the lowland ones. The larger dots represent the significant trends.}
    \label{fig:monthly_mean_temperature}
\end{figure}

 \smallskip 
Figure \ref{fig:monthly_mean_temperature} represents the monthly mean temperature trends derived from Sen's slope estimator for the 6 distinct regions: the UK highland and lowland in the upper part, the Italian Valle d'Aosta region in the middle, and the Italian Piemonte region in the lower part of the figure. On the left are the highlands, while on the right are the lowland stations. Larger, bold data points highlight statistically significant warming and cooling trends. 

As shown in Figure~\ref{fig:monthly_mean_temperature}, both negative and positive trends occurred across the months. In the UK the slopes are generally lower than in Italy. For the UK, January to April show a decrease in temperature, while the Italian stations exhibit positive slopes for most of the months. 
\par \smallskip
Going down into detail, the Italian stations generally displayed positive (warming) trends in summer, as well as in February, April, and December. 
In February, July, and August, the Italian highland stations recorded monthly temperature trends with values above  $0.15^{\circ}C / year$, while May and June showed values lower than $-0.06^{\circ}C / year$. Meanwhile, Italian lowland stations showed higher monthly temperature trends in January, February, April, July, August, and September raised above $0.06 ^{\circ}C / year$ in most of the stations with the exception of Fossano which showed negative trends in the first three months different from other stations in its group, and lower values in March and May, falling below $-0.06^{\circ}C / year$. Specifically, the trends of all the stations in the Italian lowlands and highlands show a negative trend in May.  Notably, the temperature patterns at the UK highland stations were different from those in the other two Italian groups but comparable to those in the UK lowlands (see Figure~\ref{fig:monthly_mean_temperature}). 
December stood out as the month with the highest trends at both highland and lowland stations in the UK. The temperature increased between 0.01 and $0.10^{\circ}C / year$ in the highlands and between 0.01 and $0.06^{\circ}C / year$ in the lowlands. The month of December in Italy also recorded generally quite high trends. Instead, stations in the UK behave in the opposite way to Italy in May, with an increase in trend values, while Italy everywhere shows a decrease. Another opposite behavior occurs in April, when all Italian stations showed an increasing trend, while in the UK, April is one of the months with the most significant decreasing trend, ranging between minus 0.04 and $0.12 ^{\circ}C /year$.
\par \smallskip
Taking into account statistical significance, among the noteworthy results, there are two statistically significant cooling trends: one in January at the UK highland station Bealach, and another in April at the UK lowland station Aboyne.  Furthermore, statistically significant warming trends were observed in August for the Gressoney and Valgrisenche stations, as well as in September and February for the Lillianes station. 
Saint Marcel, Avigliana, Borgone, and Luserna stations in the Italian lowland group displayed statistically significant warming trends in different months: Saint Marcel in August, Avigliana in January, April, September, and November; Borgone in August, September, and November; Luserna in July, September, and November. Specifically, between September and November, three lowland Italian stations reported statistically significant warming trends; Avigliana exhibited the highest frequency of these warming trends. 

\par \smallskip
Now we explore the resemblance of the temporal patterns of monthly mean temperature slopes (visible in Figure \ref{fig:monthly_mean_temperature}) within and across the $6$ groups. This analysis is a novelty in the literature of this field and relies on the hierarchical clustering technique in conjunction with regularized DTW. The results of hierarchical clustering for monthly mean temperatures are shown in Table \ref{tab:DTW_similar_measure}: stations in each of the $6$ groups of stations are assigned to one of the four clusters each month. The performance of this procedure is evaluated using the Silhouette score, as shown in the last column of Table \ref{tab:DTW_similar_measure}.
The results of hierarchical clustering for the slopes of monthly mean air temperatures are shown in Table \ref{tab:slope_cluster}
\par \smallskip
Prior to analyzing the monthly mean temperature and the slopes of monthly mean temperature hierarchical clustering within and across the groups, sensitivity analysis  was examined as  shown in Table~\ref{tab:silhouette_score} and determined the number of clusters. Four clusters are considered in this study. The choice of $4$ clusters is mainly motivated by the results depicted in Table~\ref{tab:silhouette_score}. Thus, the main goal is to test whether the observed data characterized the area to such an extent that they were found grouped in the same clusters. For completeness, a range of $2$ to $6$ was considered for the number of clusters, and the corresponding Silhouette Scores are shown in Table~\ref{tab:silhouette_score}.
Table~\ref{tab:silhouette_score} specifying $4$ clusters yields a higher overall mean Silhouette Score than the other settings. Reading Table~\ref{tab:silhouette_score} along the rows and comparing the scores for a given month across clusters, the $4$ clusters have the highest Silhouette score in the months from March to November.  Based on this finding, together with the analysis of the geographical location of the stations and the shape of the dendrogram diagrams generated by hierarchical clustering, the optimal number of clusters in this study is $4$.

\setcounter{table}{3}
 \begin{table}[H]
  \centering  
    \begin{tabular}{|p{6.1em}|c|c|c|c|c|}
    \toprule
    \multicolumn{1}{|c|}{} & \multicolumn{1}{c}{} & \multicolumn{1}{c}{ {Silhouette Scores}} & \multicolumn{1}{c}{} & \multicolumn{1}{c}{} &  \\
\cmidrule{2-6}    \textbf{month} & \multicolumn{1}{p{7em}|}{\textbf{$\quad 2$ clusters}} & \multicolumn{1}{p{6.445em}|}{\textbf{$\quad 3$ clusters}} & \multicolumn{1}{p{6.445em}|}{\textbf{$\quad 4$ clusters}} & \multicolumn{1}{p{6.445em}|}{\textbf{$\quad 5$ clusters}} & \multicolumn{1}{p{6.445em}|}{\textbf{$\quad 6$ clusters}} \\
    \midrule
    January & \textbf{0.64} & 0.44  & 0.40   & 0.36  & 0.32 \\
    \midrule
    February & \textbf{0.51} & 0.49  & 0.48  & 0.37  & 0.37 \\
    \midrule
    March & 0.61  & 0.63  & \textbf{0.64} & 0.55  & 0.41 \\
    \midrule
    April & 0.65  & 0.67  & \textbf{0.69} & 0.58  & 0.6 \\
    \midrule
    May   & 0.65  & 0.67  & \textbf{0.68} & 0.62  & 0.62 \\
    \midrule
    June  & 0.62  & 0.68  & \textbf{0.71} & 0.69  & 0.66 \\
    \midrule
    July  & 0.61  & 0.67  & \textbf{0.72} & 0.69  & 0.67 \\
    \midrule
    August & 0.59  & 0.66  & \textbf{0.74} & 0.73  & 0.7 \\
    \midrule
    September & 0.62  & 0.67  & \textbf{0.64} & 0.63  & 0.65 \\
    \midrule
    October & \textbf{0.63} & \textbf{0.63} & \textbf{0.63} & 0.53  & 0.47 \\
    \midrule
    November & 0.54  & 0.51  & \textbf{0.55} & 0.45  & 0.47 \\
    \midrule
    December & \textbf{0.44} & 0.42  & 0.32  & 0.24  & 0.27 \\
    \midrule
    Overall mean &  0.593 & 0.595 & \textbf{0.600} & 0.537 & 0.518 \\
    \bottomrule
    \end{tabular}
\medskip
    \caption{The Silhouette Scores of monthly mean temperature hierarchical clustering with a different number of clusters. The higher Silhouette Score over each row is highlighted in bold.}
  \label{tab:silhouette_score}%
\end{table}%

\par \smallskip
Then the hierarchical clustering is implemented using $4$ clusters.
According to the experimental results of hierarchical clustering within a group, Italian highland stations are consistently classified into two clusters (Cluster I and Cluster II), while Italian lowland stations are consistently classified into Cluster IV in all months, except January, November, and December (see the $10$ stations clustered in Cluster I and Susa, Donnas Clapey clustered in Cluster III in January and in December the four stations, Carmagnola, Marcel Surpion, Aosta Mont Fleury, Saint Christophe are clustered in Cluster I).  UK lowland stations are classified in the same cluster for all months, however, their cluster assignments differ due to the similarity measure across the $6$ groups. 
UK lowland stations are grouped into three clusters: Cluster I spans from March to May, and September to November, Cluster II spans from June to August, and Cluster IV spans from December to February. UK highland stations are consistently classified into the same cluster in seven different months and unconsistently clustered in the other five months.
\par \smallskip
The Italian highland's Valpelline, Lillianes, Coazze, and Valclarea stations, at $1029 \, m$, $1256 \, m$, $1130 \, m$ and $1068 \,m$, respectively, were identified in the same clusters for every month, suggesting comparable patterns in monthly mean temperatures.
The other four Italian highland stations, Gressoney at $1642 \, m$, Ollomont at $2017 \, m$, Valgrisenche at $1859 \, m$, and Cogne at $1682 \, m$, are likewise grouped into a unique cluster as they show a similar monthly mean temperature pattern. Consequently, the Italian highland stations are organized into two clusters except January and November, reflecting similarities in monthly average temperature patterns (see also Table \ref{tab:DTW_similar_measure}). This clustering is confirmed by the Silhouette Score with a range from $0.32$ in December to $0.74$ in August.

\par \smallskip
All the stations situated in the Italian lowland - Avigliana at $340 \, m$, Borgone at $400 \,m$, Carmagnola at $232\, m$, Luserna at $475\, m$, Susa at $470\, m$, Costigliole at $440 \,m$, Fossano at $403\, m$, Aosta Mont Fleury at $577 \,m$, Donnas Clapey at $318 \,m$, Marcel Surpion at $540 \,m$, Champdepraz Ponte Dora at $370 \, m$, and Saint Christopher at $545 \,m$ - are consistently clustered together. These stations were assigned to Cluster IV for nine months, with Silhouette Scores ranging from $0.48$ to $0.74,$ and to Cluster I, IV in the remaining months, with Silhouette Scores of  $0.55$ and $0.40$ respectively. The only exception was the Susa and Donnas Clapey stations which were always assigned to the same Cluster. The Susa and Donnas Clapey stations are grouped in Cluster III  with no intra- and inter-cluster to other stations in December.
\par \smallskip
Considering the UK highland, most months showed a Silhouette Score between $0.64$ and $0.74,$  that is, a uniform clustering pattern among all stations. All stations are grouped in Cluster III. Among the UK highland stations GreatDun and Bealach are clustered in Cluster III in January, November, and Cluster I in December while Aonach Mor and Cairngorm are clustered in Cluster III in October.  In general, in the UK highlands, the monthly mean temperature showed uniform patterns in most months. 

\par \smallskip
From the monthly mean temperature patterns across different groups, the UK lowland stations, the Italian lowland stations, and the UK highland stations have distinct clusters for every month, corresponding to distinct patterns.
The four Italian highland stations are grouped consistently in Cluster II with the UK lowland stations in June, July, and August.  It also clustered with some stations of the UK highland stations in January, February, October, November, and December.


The UK lowland stations showed a similar pattern to the Italian highland, with Valpelline, Lillianes, Valclarea, and Coazze assigned to Cluster I consistently and the other four stations assigned to Cluster II. In addition, the Valpelline, Lillianes, Valclarea, and Coazze stations consistently show patterns similar to some of the Italian lowland stations in January, November, and December.
Among all the stations, there is no consistent pattern shared between the Italian and UK highland stations. However, in some months, four of the Italian highland stations exhibited a pattern similar to that of the UK highland stations. Furthermore, the Italian lowland stations and the UK highland stations did not display consistent monthly mean temperature patterns for all months. 
Consequently, it can be concluded that monthly mean temperature patterns do not exhibit any persistent similarities between groups and that each group continues to exhibit its unique and stable monthly mean temperature features. However, UK lowlands and Italian highlands showed some sort of similarity in Cluster I and Cluster II for most months. 

\par \smallskip

\setcounter{table}{5}
\begin{table}[H]
  \centering
  \resizebox{18cm}{10.5cm}{
  \begin{tabular}{|c|l|l|l|l|l|l|l|}
    \toprule
    \multicolumn{1}{|c|}{\textbf{\footnotesize Month}} & \multicolumn{1}{c|}{\textbf{ \footnotesize Group}} & \textbf{\footnotesize Region} & \multicolumn{1}{c|}{\textbf{\footnotesize {Cluster I}} }& \multicolumn{1}{c|}{\textbf{\footnotesize{ Cluster II}}} & \multicolumn{1}{c|}{\textbf{\footnotesize{ Cluster III}}} & \multicolumn{1}{c|}{\textbf{\footnotesize{ Cluster IV}} }& \bf{\footnotesize S.Score} \\
    \midrule
    
\multirow{6}[12]{*}{\footnotesize January} &
\multicolumn{1}{l|}{\multirow{2}[4]{*}\bf{{\footnotesize IH}}} & \bf{\footnotesize {Valle d'Aosta}} &  \small{VP, LL}     &   \small{OL, VG, CG, GR}    &       &  & \\
 
& \multicolumn{1}{l|}{} & \bf{\footnotesize Piemonte} &  \small{CO, VC}     &       &  \footnotesize{BA, NI, PR}     &  \\

\cmidrule{2-7}        
& \multicolumn{1}{l|}{\multirow{2}[4]{*}{\bf{\footnotesize IL}}} & \bf{\footnotesize{Valle d'Aosta}} & \small{SC, CP, MF, MS}      &       &       & \small{DC}& \small{0.40}\\
       
& \multicolumn{1}{l|}{} & \bf{\footnotesize{Piemonte}} & \small{CM, AG, BG, CT, LS, FS }     &       &       & \footnotesize{SS} & \\

\cmidrule{2-7}          
& \multicolumn{2}{l|}{\bf{\footnotesize{UKH}}} &       &  \small{CR, AN, CW}    &  \footnotesize{BL, GD}     & & \\
\cmidrule{2-7}          
& \multicolumn{2}{l|}{\bf{\footnotesize{UKL}}} &       &       &       & \small{WR, TH, AY, AR}
& \\
    \bottomrule

\multirow{6}[12]{*}{\footnotesize February} &
\multicolumn{1}{l|}{\multirow{2}[4]{*}\bf{{\footnotesize IH}}} & \bf{\footnotesize {Valle d'Aosta}} &  \small{VP, LL }     &   \small{OL, VG, CG, GR}    &       &  & \\
 
& \multicolumn{1}{l|}{} & \bf{\footnotesize Piemonte} &  \small{ CO, VC, BA, NI, PR}     & \footnotesize{}      &       &  \\

\cmidrule{2-7}        
& \multicolumn{1}{l|}{\multirow{2}[4]{*}{\bf{\footnotesize IL}}} & \bf{\footnotesize{Valle d'Aosta}} & \footnotesize{}      &       &       & \small{SC, CP, MF, DC, MS }& \footnotesize{0.48}\\
       
& \multicolumn{1}{l|}{} & \bf{\footnotesize{Piemonte}} & \small{ }     &       &       & \footnotesize{CM, AG, BG, CT, LS, FS, SS } & \\
\cmidrule{2-7}          
& \multicolumn{2}{l|}{\bf{\footnotesize{UKH}}} &       &  \small{CR , AN}    & \small{BL, GD, CW} \footnotesize{}     & & \\
\cmidrule{2-7}          
& \multicolumn{2}{l|}{\bf{\footnotesize{UKL}}} &       &       &       & \small{WR, TH, AY, AR}
& \\
    \bottomrule
   
\multirow{6}[12]{*}{\footnotesize March} &
\multicolumn{1}{l|}{\multirow{2}[4]{*}\bf{{\footnotesize IH}}} & \bf{\footnotesize {Valle d'Aosta}} &  \small{VP, LL }     &   \small{OL, VG, CG, GR}    &       &  & \\
 
& \multicolumn{1}{l|}{} & \bf{\footnotesize Piemonte} &  \small{ CO, VC, BA, NI, PR}     & \footnotesize{}      &       &  \\

\cmidrule{2-7}        
& \multicolumn{1}{l|}{\multirow{2}[4]{*}{\bf{\footnotesize IL}}} & \bf{\footnotesize{Valle d'Aosta}} & \footnotesize{}      &       &       & \small{SC, CP, MF, DC, MS }& \footnotesize{0.64}\\
       
& \multicolumn{1}{l|}{} & \bf{\footnotesize{Piemonte}} & \small{ }     &       &       & \footnotesize{CM, AG, BG, CT, LS, FS, SS } & \\
\cmidrule{2-7}          
& \multicolumn{2}{l|}{\bf{\footnotesize{UKH}}} &       &  \small{}    & \small{CR , AN, BL, GD, CW} \footnotesize{}     & & \\
\cmidrule{2-7}          
& \multicolumn{2}{l|}{\bf{\footnotesize{UKL}}} &  \small{WR, TH, AY, AR}     &       &       & 
& \\
    \bottomrule

\multirow{6}[12]{*}{\footnotesize April} &
\multicolumn{1}{l|}{\multirow{2}[4]{*}\bf{{\footnotesize IH}}} & \bf{\footnotesize {Valle d'Aosta}} &  \small{VP, LL }     &   \small{OL, VG, CG, GR}    &       &  & \\
 
& \multicolumn{1}{l|}{} & \bf{\footnotesize Piemonte} &  \small{ CO, VC, BA, NI, PR}     & \footnotesize{}      &       &  \\

\cmidrule{2-7}        
& \multicolumn{1}{l|}{\multirow{2}[4]{*}{\bf{\footnotesize IL}}} & \bf{\footnotesize{Valle d'Aosta}} & \footnotesize{}      &       &       & \small{SC, CP, MF, DC, MS }& \footnotesize{0.69}\\
       
& \multicolumn{1}{l|}{} & \bf{\footnotesize{Piemonte}} & \small{ }     &       &       & \footnotesize{CM, AG, BG, CT, LS, FS, SS } & \\
\cmidrule{2-7}          
& \multicolumn{2}{l|}{\bf{\footnotesize{UKH}}} &       &  \small{}    & \small{CR , AN, BL, GD, CW} \footnotesize{}     & & \\
\cmidrule{2-7}          
& \multicolumn{2}{l|}{\bf{\footnotesize{UKL}}} &  \small{WR, TH, AY, AR}     &       &       & 
& \\
    \bottomrule

\multirow{6}[12]{*}{\footnotesize May} &
\multicolumn{1}{l|}{\multirow{2}[4]{*}\bf{{\footnotesize IH}}} & \bf{\footnotesize {Valle d'Aosta}} &  \small{VP, LL }     &   \small{OL, VG, CG, GR}    &       &  & \\
 
& \multicolumn{1}{l|}{} & \bf{\footnotesize Piemonte} &  \small{ CO, VC, BA, NI, PR}     & \footnotesize{}      &       &  \\

\cmidrule{2-7}        
& \multicolumn{1}{l|}{\multirow{2}[4]{*}{\bf{\footnotesize IL}}} & \bf{\footnotesize{Valle d'Aosta}} & \footnotesize{}      &       &       & \small{SC, CP, MF, DC, MS }& \footnotesize{0.68}\\
       
& \multicolumn{1}{l|}{} & \bf{\footnotesize{Piemonte}} & \small{ }     &       &       & \footnotesize{CM, AG, BG, CT, LS, FS, SS } & \\
\cmidrule{2-7}          
& \multicolumn{2}{l|}{\bf{\footnotesize{UKH}}} &       &  \small{}    & \small{CR , AN, BL, GD, CW} \footnotesize{}     & & \\
\cmidrule{2-7}          
& \multicolumn{2}{l|}{\bf{\footnotesize{UKL}}} &  \small{WR, TH, AY, AR}     &       &       & 
& \\
    \bottomrule

\multirow{6}[12]{*}{\footnotesize June} &
\multicolumn{1}{l|}{\multirow{2}[4]{*}\bf{{\footnotesize IH}}} & \bf{\footnotesize {Valle d'Aosta}} &  \small{VP, LL }     &   \small{OL, VG, CG, GR}    &       &  & \\
 
& \multicolumn{1}{l|}{} & \bf{\footnotesize Piemonte} &  \small{ CO, VC, BA, NI, PR}     & \footnotesize{}      &       &  \\

\cmidrule{2-7}        
& \multicolumn{1}{l|}{\multirow{2}[4]{*}{\bf{\footnotesize IL}}} & \bf{\footnotesize{Valle d'Aosta}} & \footnotesize{}      &       &       & \small{SC, CP, MF, DC, MS }& \footnotesize{0.71}\\
       
& \multicolumn{1}{l|}{} & \bf{\footnotesize{Piemonte}} & \small{ }     &       &       & \footnotesize{CM, AG, BG, CT, LS, FS, SS } & \\
\cmidrule{2-7}          
& \multicolumn{2}{l|}{\bf{\footnotesize{UKH}}} &       &  \small{}    & \small{CR , AN, BL, GD, CW} \footnotesize{}     & & \\
\cmidrule{2-7}          
& \multicolumn{2}{l|}{\bf{\footnotesize{UKL}}} &       & \small{WR, TH, AY, AR}      &       & 
& \\
    \bottomrule

    \end{tabular}%
    }
    \vspace{0.1cm}
     
\end{table}%

\setcounter{table}{5}
\begin{table}[H]
\ContinuedFloat
  \centering
  \resizebox{18cm}{10cm}{
  \begin{tabular}{|c|l|l|l|l|l|l|l|}
    \toprule
    \multicolumn{1}{|c|}{\textbf{\footnotesize Month}} & \multicolumn{1}{c|}{\textbf{ \footnotesize Group}} & \textbf{\footnotesize Region} & \multicolumn{1}{c|}{\textbf{\footnotesize {Cluster I}} }& \multicolumn{1}{c|}{\textbf{\footnotesize{ Cluster II}}} & \multicolumn{1}{c|}{\textbf{\footnotesize{ Cluster III}}} & \multicolumn{1}{c|}{\textbf{\footnotesize{ Cluster IV}} }& \bf{\footnotesize S.Score} \\
    \midrule

\multirow{6}[12]{*}{\footnotesize July} &
\multicolumn{1}{l|}{\multirow{2}[4]{*}\bf{{\footnotesize IH}}} & \bf{\footnotesize {Valle d'Aosta}} &  \small{VP, LL }     &   \small{OL, VG, CG, GR}    &       &  & \\
 
& \multicolumn{1}{l|}{} & \bf{\footnotesize Piemonte} &  \small{ CO, VC, BA, NI, PR}     & \footnotesize{}      &       &  \\

\cmidrule{2-7}        
& \multicolumn{1}{l|}{\multirow{2}[4]{*}{\bf{\footnotesize IL}}} & \bf{\footnotesize{Valle d'Aosta}} & \footnotesize{}      &       &       & \small{SC, CP, MF, DC, MS }& \footnotesize{0.72}\\
       
& \multicolumn{1}{l|}{} & \bf{\footnotesize{Piemonte}} & \small{ }     &       &       & \footnotesize{CM, AG, BG, CT, LS, FS, SS } & \\
\cmidrule{2-7}          
& \multicolumn{2}{l|}{\bf{\footnotesize{UKH}}} &       &  \small{}    & \small{CR , AN, BL, GD, CW} \footnotesize{}     & & \\
\cmidrule{2-7}          
& \multicolumn{2}{l|}{\bf{\footnotesize{UKL}}} &       & \small{WR, TH, AY, AR}      &       & 
& \\
    \bottomrule

\multirow{6}[12]{*}{\footnotesize August} &
\multicolumn{1}{l|}{\multirow{2}[4]{*}\bf{{\footnotesize IH}}} & \bf{\footnotesize {Valle d'Aosta}} &  \small{VP, LL }     &   \small{OL, VG, CG, GR}    &       &  & \\
 
& \multicolumn{1}{l|}{} & \bf{\footnotesize Piemonte} &  \small{ CO, VC, BA, NI, PR}     & \footnotesize{}      &       &  \\

\cmidrule{2-7}        
& \multicolumn{1}{l|}{\multirow{2}[4]{*}{\bf{\footnotesize IL}}} & \bf{\footnotesize{Valle d'Aosta}} & \footnotesize{}      &       &       & \small{SC, CP, MF, DC, MS }& \footnotesize{0.74}\\
       
& \multicolumn{1}{l|}{} & \bf{\footnotesize{Piemonte}} & \small{ }     &       &       & \footnotesize{CM, AG, BG, CT, LS, FS, SS } & \\
\cmidrule{2-7}          
& \multicolumn{2}{l|}{\bf{\footnotesize{UKH}}} &       &  \small{}    & \small{CR , AN, BL, GD, CW} \footnotesize{}     & & \\
\cmidrule{2-7}          
& \multicolumn{2}{l|}{\bf{\footnotesize{UKL}}} &       & \small{WR, TH, AY, AR}      &       & 
& \\
    \bottomrule

\multirow{6}[12]{*}{\footnotesize September} &
\multicolumn{1}{l|}{\multirow{2}[4]{*}\bf{{\footnotesize IH}}} & \bf{\footnotesize {Valle d'Aosta}} &  \small{VP, LL }     &   \small{OL, VG, CG, GR}    &       &  & \\
 
& \multicolumn{1}{l|}{} & \bf{\footnotesize Piemonte} &  \small{ CO, VC, BA, NI, PR}     & \footnotesize{}      &       &  \\

\cmidrule{2-7}        
& \multicolumn{1}{l|}{\multirow{2}[4]{*}{\bf{\footnotesize IL}}} & \bf{\footnotesize{Valle d'Aosta}} & \footnotesize{}      &       &       & \small{SC, CP, MF, DC, MS }& \footnotesize{0.64}\\
       
& \multicolumn{1}{l|}{} & \bf{\footnotesize{Piemonte}} & \small{ }     &       &       & \footnotesize{CM, AG, BG, CT, LS, FS, SS } & \\
\cmidrule{2-7}          
& \multicolumn{2}{l|}{\bf{\footnotesize{UKH}}} &       &  \small{}    & \small{CR , AN, BL, GD, CW} \footnotesize{}     & & \\
\cmidrule{2-7}          
& \multicolumn{2}{l|}{\bf{\footnotesize{UKL}}} & \small{WR, TH, AY, AR}       &      &       & 
& \\
    \bottomrule

\multirow{6}[12]{*}{\footnotesize October} &
\multicolumn{1}{l|}{\multirow{2}[4]{*}\bf{{\footnotesize IH}}} & \bf{\footnotesize {Valle d'Aosta}} &  \small{VP, LL }     &   \small{OL, VG, CG, GR}    &       &  & \\
 
& \multicolumn{1}{l|}{} & \bf{\footnotesize Piemonte} &  \small{ CO, VC, BA, NI, PR}     & \footnotesize{}      &       &  \\

\cmidrule{2-7}        
& \multicolumn{1}{l|}{\multirow{2}[4]{*}{\bf{\footnotesize IL}}} & \bf{\footnotesize{Valle d'Aosta}} & \footnotesize{}      &       &       & \small{SC, CP, MF, DC, MS }& \footnotesize{0.63}\\
       
& \multicolumn{1}{l|}{} & \bf{\footnotesize{Piemonte}} & \small{ }     &       &       & \footnotesize{CM, AG, BG, CT, LS, FS, SS } & \\
\cmidrule{2-7}          
& \multicolumn{2}{l|}{\bf{\footnotesize{UKH}}} &       &  \small{BL, GD, CW}    & \small{CR , AN } \footnotesize{}     & & \\
\cmidrule{2-7}          
& \multicolumn{2}{l|}{\bf{\footnotesize{UKL}}} & \small{WR, TH, AY, AR}       &      &       & 
& \\
    \bottomrule

\multirow{6}[12]{*}{\footnotesize November} &
\multicolumn{1}{l|}{\multirow{2}[4]{*}\bf{{\footnotesize IH}}} & \bf{\footnotesize {Valle d'Aosta}} &  \small{VP, LL }     &   \small{OL, VG, CG, GR}    &       &  & \\
 
& \multicolumn{1}{l|}{} & \bf{\footnotesize Piemonte} &  \small{ CO, VC}  &      &  \small{BA, NI, PR}     & & \\

\cmidrule{2-7}        
& \multicolumn{1}{l|}{\multirow{2}[4]{*}{\bf{\footnotesize IL}}} & \bf{\footnotesize{Valle d'Aosta}} & \footnotesize{}      & SC, MF, MS     &       & \small{ CP, DC }& \footnotesize{0.55}\\
       
& \multicolumn{1}{l|}{} & \bf{\footnotesize{Piemonte}} & \small{ }     &       &       & \footnotesize{CM, AG, BG, CT, LS, FS, SS } & \\
\cmidrule{2-7}          
& \multicolumn{2}{l|}{\bf{\footnotesize{UKH}}} &       &  \small{CR , AN, CW}    & \small{BL, GD } \footnotesize{}     & & \\
\cmidrule{2-7}          
& \multicolumn{2}{l|}{\bf{\footnotesize{UKL}}} & \small{WR, TH, AY, AR}       &   & & & \\
    \bottomrule

\multirow{6}[12]{*}{\footnotesize December} &
\multicolumn{1}{l|}{\multirow{2}[4]{*}\bf{{\footnotesize IH}}} & \bf{\footnotesize {Valle d'Aosta}} &  \small{VP, LL }     &   \small{OL, VG, CG, GR}    &       &  & \\
 
& \multicolumn{1}{l|}{} & \bf{\footnotesize Piemonte} &  \small{ CO, VC}  &      &  \small{BA, NI, PR}     & & \\

\cmidrule{2-7}        
& \multicolumn{1}{l|}{\multirow{2}[4]{*}{\bf{\footnotesize IL}}} & \bf{\footnotesize{Valle d'Aosta}} & \small{SC, MF, MS }  &     &  \small{DC}     & \small{ CP }& \footnotesize{0.32}\\
       
& \multicolumn{1}{l|}{} & \bf{\footnotesize{Piemonte}} & \small{CM }     &       &  \small{SS}     & \footnotesize{AG, BG, CT, LS, FS } & \\
\cmidrule{2-7}          
& \multicolumn{2}{l|}{\bf{\footnotesize{UKH}}} &  \small{BL, GD}     &  \small{CR , AN, CW}    &   & & \\
\cmidrule{2-7}          
& \multicolumn{2}{l|}{\bf{\footnotesize{UKL}}} &        &      &       & \small{WR, TH, AY, AR}& \\
    \bottomrule
\end{tabular}%

    }
   \caption{\footnotesize{The classification of the $32$ stations into $4$ clusters using hierarchical clustering in conjunction with regularized DTW. The values of the Silhouette Score are given in the last column. }}
\label{tab:DTW_similar_measure}
\end{table}%

\par \smallskip
Table~\ref{tab:slope_cluster} shows the results of the hierarchical clustering with the DTW applied to the slopes of the monthly mean air temperature. Despite the obvious distinction  
between positive and negative slopes, Table~\ref{tab:slope_cluster} adds some further insights into understanding similarities/dissimilarities among stations although the estimated slopes are rather uncertain. Indeed all of the stations in the UK highland and UK lowland are grouped into Cluster IV with the exception of Aonach Mor station in the UK highland grouped at Cluster I. All of the Italian lowland stations are grouped into Cluster III. However, Fossano station is grouped in Cluster IV together with UK lowland and UK highland stations. The slope of the Italian highland station Valpelline is clustered in Cluster I together with the UK highland station Aonach Mor, whereas the Italian lowland station Fossano is clustered together with the UK highland and the UK lowland stations in Cluster IV.
These unusual behaviors with respect to their belonging group can also be observed in Figure~\ref{fig:monthly_mean_temperature}. Indeed Valpelline in the Italian highland, Fossano in the Italian lowland, and Aonach Mor in the UK highland have distinct slope patterns within their reference groups.  In particular, the slopes of the two Italian stations often trend lower than the lower slopes of their reference group. The opposite happens for the slope of the Aonach Mor. 
The fact that these two stations were grouped in Cluster I, thus showing different behaviors from all the others, may be due to the following reasons: Valpelline is the lowest of the mountain stations in Valle d'Aosta - and in fact its slopes in Figure 4 are different from the others - while Aonach Mor has 23$\%$  missing values. The Fossano  station is the southernmost station in the entire Italian set and is beginning to have some sublittoral climatic effects.
It is worth noting that this difference does not appear in the average data, and thus in Table 5, but its trends in Figure 4, especially in winter, are very different from those of the other stations. Other differences in the values of slopes related to specific months seem not to affect the clustering. For example, the Cairngorm station showed a different pattern in two months (June and July). However, the hierarchical clustering tolerated this deviation and clustered Cairngorm along with his group. In conclusion, we can infer that the slopes of the UK highland and lowland stations are consistently grouped in Cluster IV, with the exception of Aonach Mor.  The slopes of Italian highland and lowland stations are grouped in Cluster II and Cluster III, respectively, with the exception of Valpelline and Fossano.

\setcounter{table}{5}
\begin{table}[H]
  \centering
   \begin{tabular}{|ll|r|r|r|l|}
    \toprule
    \multicolumn{1}{|l|}{\textbf{Group}} & \textbf{Region} & \multicolumn{1}{l|}{\textbf{Cluster I}} & \multicolumn{1}{l|}{\textbf{Cluster II}} & \multicolumn{1}{l|}{\textbf{Cluster III}} & \textbf{Cluster IV} \\
    \midrule
    \multicolumn{1}{|c|}{\multirow{2}[4]{*}{\textbf{IH}}} & \textbf{Valle d'Aosta} & \multicolumn{1}{l|}{VP} & \multicolumn{1}{l|}{LL, OL, VG, CG, GR} &       &  \\
\cmidrule{2-6}    \multicolumn{1}{|c|}{} & \textbf{Piemonte} &       & \multicolumn{1}{l|}{CO, VC, BA, NI, PR} &       &  \\
    \midrule
    \multicolumn{1}{|c|}{\multirow{2}[4]{*}{\textbf{IL}}} & \textbf{Valle d'Aosta} &       &       & \multicolumn{1}{l|}{SC,  CP,  MF, DC, MS} &  \\
\cmidrule{2-6}    \multicolumn{1}{|c|}{} & \textbf{Piemonte} &       &       & \multicolumn{1}{l|}{CM, AG, BG, CT, LS, SS} & FS \\
    \midrule
    \multicolumn{2}{|l|}{\textbf{UKH}} & \multicolumn{1}{l|}{AN} &       &       & CR, BL, GD, CW  \\
    \midrule
    \multicolumn{2}{|l|}{\textbf{UKL}} &       &       &       &  WR, TH, AY,  AR \\
    \bottomrule
    \end{tabular}%
    \vspace{0.2cm}
 \caption{The classification of the trends of the $32$ stations into $4$ clusters using hierarchical clustering in conjunction with DTW}
    \label{tab:slope_cluster}
\end{table}%

\begin{figure}[H]
    \centering
    \includegraphics[width=0.9\textwidth, height=1.12\textwidth]{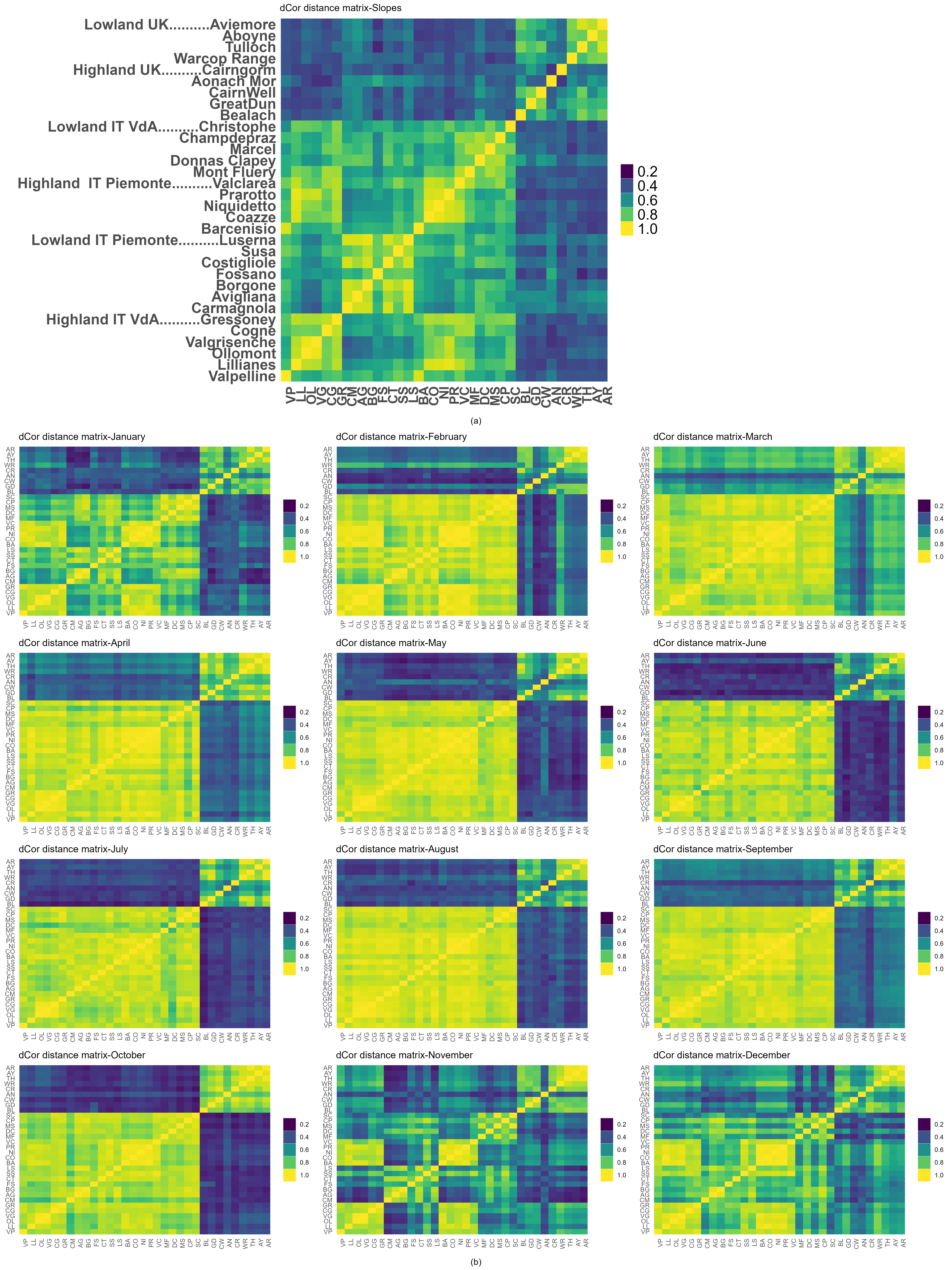}
    \caption{Distance correlation matrix of the 32 stations across Italy} (highland and lowland) and the UK (highland and lowland). (a) matrix for the slopes of the monthly mean, and (b) the monthly mean temperature.
    \label{fig:commonlabel}
\end{figure}

The distance correlation  (see Figure~\ref{fig:commonlabel} (a)) of the $32$ stations is employed to quantify the strength of relationships of the trends of the monthly mean temperature. The stations in the highland of  Piemonte showed a relatively strong correlation with the highland of Valle d'Aosta. We can see that the station has a relatively strong correlation within the $6$ groups and a relatively weaker correlation between different groups, especially between Italian and UK stations. 

\par \smallskip
Finally, the distance correlation matrix of the $32$ stations in each month is presented in Figure~\ref{fig:commonlabel} (b). Interestingly, there was a strong correlation between every station and other stations in the same group. 
The Piemonte and Valle d'Aosta highland stations are highly correlated in the monthly mean temperature for every month {(see Figure~\ref{fig:commonlabel} (b))}. It is also evident that the Italian highland and lowland stations are highly correlated in the monthly mean temperature for every month, with the exception of November and December. Most of the stations in the UK lowland and highland areas showed relatively strong correlations with each other in the monthly mean temperature for every month. However, the UK stations showed a weak correlation with the Italian stations, as depicted in Figure~\ref{fig:commonlabel} (b). In general, the Italian lowland stations exhibited a weaker correlation with the Italian highland stations in January, and for some stations, this also occurred in November and December. Overall, the Italian highland and lowland stations show a stronger correlation.
In comparison with Switzerland \citep{rebetez2008monthly}, the trends of Italian highland and lowland show less spreading in monthly trends, perhaps due to the fact that the Switzerland analysis was done in the period 1975-2004. Also \cite{bruley2022enhanced} didn't find a difference with altitude in monthly trends in the period from 1980 to 2015 in the Massif Central of France.
 Natural climate variability, however, poses inherent limits to climate predictability which vary between areas with relatively lower or higher climate variability. The findings of our study can help to obtain a clearer picture of the time patterns, showing how they vary in different regions of Europe and at different altitudes. Understanding why the different months of the year behave in different way should require a more detailed study, but our results can provide the starting point for it, suggesting the employment of DTW and clustering to extend the  analysis in many different areas of the world. For example in the Mongolian plateau between 1986 and 2004, an exceptional warming occurred, boosted by internal variability \citep{cai2024recent}. In our results, the Italian highland trends are not enhanced with respect to lowland stations. Also ~\citet{rogora2004recent} didn't find a relation with the altitude in North-West Italian data, while ~\citet{acquaotta2015temperature} found it in the same Italian area. Indeed, even if all Italian stations in the present work show a general coherence of most monthly trends, in Figure ~\ref{fig:monthly_mean_temperature} it is visible a better correlation from November to February of mountain stations in both Italian areas, even if they are localized at about 70 $km$ distance. On the map, it is easy to see Valle d'Aosta to the North and Borgone to the South. This result can also avoid doubts about the possible errors in measuring air temperature over snow, as proved by Huwald et al. (2009). In fact the number of days with snow in the different stations can largely vary, because of altitudes that are spread between 1000 and 2000$m$ asl. Also, \cite{salerno2023local} recently found an unpredicted cooling in an area of high mountains in Himalaya. The relevant warming trends in summer in the Italian stations, instead, can have important effects on the vegetation and carbon feedback \citep{zhang2022future}. 

Regarding the possible causes of the monthly patterns, it is possible that the relatively lower warming of the UK is related to the decline of the strength of the Atlantic Meridional Overturning Circulation (AMOC) \citep{robson2016reversal, johnson2020warming}. For other insight, we refer to specific studies about the influence of dynamical drivers on monthly trends \citep{hoffmann2021identification}. Some hints for attributing trends to synoptic circulation are also in the literature, with specific reference to Europe data \citep{fleig2015attribution}. It was the first attempt to understand the influence of global change on monthly trends. As a possible progress of this work, it could be interesting to compare cloud cover and sunshine trends with temperature trends. Some interesting work has been done in Italy by \cite{manara2023variability}.

 \section{Conclusions}
 
In this paper, annual and monthly temperature trends of $32$ stations of Italian highland ($5$ stations in Piemonte and $6$ stations in Valle d'Aosta), Italian lowland ($5$ stations in Valle d'Aosta and $7$ stations in Piemonte), the UK highland ($5$ stations), and the UK lowland ($4$ stations) were analyzed, using the data collected from $2002$ to $2021$. The first purpose of the study was to analyze annual and monthly mean temperature trends. Furthermore, the unsupervised machine learning approach (hierarchical clustering in combination with DTW) is used to investigate the monthly mean air temperature patterns in order to measure the degree of similarity within and between the $6$ groups of stations. The Silhouette Score is used to assess how effectively the clustering procedure performs.
The main novelty of the paper is to show that stations having similar locations and altitudes have similar monthly slopes by quantifying them using DTW and clustering methods. These results reveal the nonrandomness of different trends along the year and among different parts of Europe, with a modest influence of altitude in wintertime. Two different regions of Europe were chosen because of the different climate and temperature trends, namely the Northern UK (smaller trends) and the North-West Italian Alps (greater trends).
\par \smallskip

The results of this study indicated a general warming trend in annual mean air temperature, with statistically significant warming observed at $8$ of the $11$ stations of the Italian highland and $9$ of the $12$ stations in the Italian lowland. Nonsignificant decreasing trends are
detected at 
 Valpelline, in the Italian highland group,
 as well as 
Fossano, in the Italian lowland group.
Conversely, the mean annual air temperature in the UK highland and the UK lowland at all stations showed a statistically nonsignificant cooling trend. In most stations, the results obtained from the parametric and nonparametric methods used in this study are comparable. The bias of distribution and data outliers on the OLS method is evident at some stations, particularly those that differ from normal ones, such as Valpelline and Aonach Mor. Due to the non-normal nature of the annual mean air temperature at Valpelline, Costigliole, Aonach Mor, and all stations in the UK lowland, differences in the magnitude of slopes and $R^2$ values are seen among the UK lowland stations when comparing the OLS and other methods. Nevertheless, all methods indicate a nonsignificant cooling trend across all UK stations.

\par \smallskip

Analyzing trends in monthly average air temperature, negative slopes were observed in May and June at most Italian stations, indicating a cooling trend. The months of February, August, and December, on the other hand, demonstrated clear warming trends. In the Italian highlands, the Valpelline station is an exception, with a decreasing trend in March, October, and November. Compared with Italian stations, the UK highland and lowland stations generally have more months with cooling trends.

\par \smallskip
Hierarchical clustering showed that stations within the same group  had similar monthly mean temperature patterns. The similarity which can be seen in Figure~\ref{fig:monthly_mean_temperature} has been justified with clustering and correlation methods as shown also in Figure~\ref{fig:commonlabel}. As an exception, the Italian highland stations are grouped into two clusters: Valpelline, Lillianes, Coazze, and Valclarea are grouped in one cluster and the other four (Ollomont, Valgrisenche, Gressoney, and Cogne) stations are grouped in another cluster in all months. The peculiarities of Valpelline, Fossano, and Aonoch Mor can be attributed respectively to the fact that Valpelline is the lowest in elevation of its group, Fossano is the most southern of the Italian ones, with some sublittoral influence, and Aonoch Mr has a large amount of missing values. It would be reasonable to increase the number of clusters by increasing the number of involved geographic areas: for example, adding midland areas. This is no doubt the subject of future research.

\par \smallskip

The main result of the paper is to clearly show the different time patterns in each month for each group of stations. This was also done using the distance correlation matrix, that shows strong correlations among the Piemonte and Valle d'Aosta highland stations every month. This pattern is also evident for most of the UK highland and lowland stations. 
These results are in agreement with the geographic location of the stations and are not too surprising. Considering a finer temporal scale, such as
daily mean temperatures, would be useful for a more comprehensive analysis of the dependence. This analysis is in progress and will be the subject of future studies. Actually, already stratifying monthly temperatures allow us to add some non-obvious observations about correlations between stations.
The Italian highland and lowland stations show a higher correlation every month with the exception of January, November, and December. Hence, the Italian lowland stations show a weaker correlation with the Italian highland stations in those three months. The distance correlation matrix depicts weak correlations among the UK and Italian stations. 

\par \smallskip
Whatever their correlation, the processes underlying the various combined processes that cause these monthly and annual trends are beyond the scope of this paper. The findings of the present paper enhance the need to understand the temperature dynamics in the different groups and altitudes of Europe. These results also emphasize the importance of continuous monitoring and analysis of data in order to better quantify climate change.

 \newpage
\dataavailability{Temperature data are freely available from MetOffice, CFVDA and ARPAPiemonte. The results of this paper are available upon request from the authors.} 
\par \smallskip
\authorcontribution{The authors confirm contribution to the paper as follows: study conception and
design: C.M.L., E.D.N., R.M., and S.F.; data curation: S.F. and C.M.L.; software programs: C.M.L.; Methodology: C.M.L. and E.D.N; Analysis and Interpretation of results: C.M.L., S.F. and E.D.N.; draft manuscript preparation: C.M.L., E.D.N., and S.F.; review and editing: C.M.L., E.D.N., R.M., and S.F.; financial sourcing: S.F.  All authors read and approved the final version of the manuscript.} 
\par \smallskip
\competinginterests{The authors declare no conflict of interest.} 
\par \smallskip
\begin{acknowledgement}
    The authors would like to acknowledge the funder of this paper. This publication is part of the project NODES which has received funding from the MUR – M4C2 1.5 of PNRR funded by the European Union - NextGenerationEU (Grant agreement no. ECS00000036). It was also partially funded by the PRIN 2022 project Snow Droughts Prediction in the Alps: A Changing Climate. We extend our heartfelt appreciation to the diligent efforts of the reviewer and the guidance of the chief editor, whose invaluable contributions have significantly enriched the quality and depth of this manuscript.
\end{acknowledgement}

\bibliographystyle{copernicus.bst}
\bibliography{mybibfile}

\end{document}